\documentclass[prb,twocolumn,showpacs,preprintnumbers,amsmath,amssymb]{revtex4}
\usepackage{graphicx}
\usepackage{color} %example: {\color{red} text}
\usepackage{amsmath}
\begin{document}

\title{Nonlinear susceptibility of a quantum 
spin glass under uniform transverse and random 
longitudinal magnetic fields}

\author{S. G. Magalhaes$^1$, C. V. Morais$^2$, F. M. Zimmer$^3$, 
M. J. Lazo$^4$, F.~D. Nobre$^5$}

\affiliation{$^1$Instituto de Fisica, Universidade Federal 
do Rio Grande do Sul, 91501-970
Porto Alegre, RS, Brazil}
\email{sgmagal@gmail.com}

\affiliation{$^3$Instituto de F\'{i}sica e Matem\'atica, Universidade 
Federal de Pelotas, 96010-900  Pelotas, RS, Brazil}

\affiliation{$^2$Departamento de Fisica, Universidade Federal de Santa Maria,
97105-900 Santa Maria, RS, Brazil}%

\affiliation{$^4$Programa de P\'os-Gradua\c{c}\~{a}o em F\'{\i}sica - Instituto de Matem\'atica, Estat\'{\i}stica e F\'{\i}sica, 
Universidade Federal do Rio Grande, 
96.201-900, Rio Grande, RS, Brazil }

\affiliation{$^5$ Centro Brasileiro de Pesquisas F\'{\i}sicas  and 
National Institute of Science and Technology for Complex Systems, 
Rua Xavier Sigaud 150, 
22290-180, Rio de Janeiro, RJ, Brazil}

\date{\today}

\begin{abstract}

The interplay between quantum fluctuations and disorder is investigated
in a quantum spin-glass model, in the presence of a uniform transverse
field $\Gamma$, as well as of a longitudinal random field $h_{i}$, which follows 
a Gaussian distribution characterized by a width proportional to $\Delta$. 
The interactions are infinite-ranged, and the model 
is studied through the replica formalism, within a one-step 
replica-symmetry-breaking procedure; 
in addition, the dependence of the Almeida-Thouless eigenvalue 
$\lambda_{\rm AT}$ (replicon) on the applied fields is analyzed. 
This study is motivated by experimental investigations on the  
LiHo$_x$Y$_{1-x}$F$_4$ compound, where   
the application of a transverse magnetic field yields
rather intriguing effects, particularly related to the behavior
of the nonlinear magnetic susceptibility $\chi_{3}$, which have 
led to a considerable experimental and theoretical debate.
We have analyzed two physically distinct situations, namely,
$\Delta$ and $\Gamma$ considered as independent, as well as 
these two quantities related, as proposed recently by some authors. 
In both cases, a spin-glass phase transition is found at a
temperature $T_{f}$, with such phase being characterized by a 
nontrivial ergodicity breaking; moreover, $T_{f}$ decreases
by increasing $\Gamma$ towards a quantum critical point at
zero temperature. 
The situation where $\Delta$ and $\Gamma$ are related 
[$\Delta \equiv \Delta(\Gamma)$] appears
to reproduce better the experimental observations on the 
LiHo$_x$Y$_{1-x}$F$_4$ compound, with the theoretical results
coinciding qualitatively 
with measurements of the nonlinear susceptibility $\chi_{3}$.
In this later case, by increasing $\Gamma$ gradually, $\chi_{3}$
becomes progressively rounded, presenting a maximum
at a temperature $T^{*}$ ($T^{*}>T_{f}$), with both 
the amplitude of the maximum and the value of $T^{*}$
decreasing gradually.   
Moreover, we also show that the random field is the main responsible 
for the smearing of the nonlinear susceptibility, acting significantly  
inside the paramagnetic phase, leading to two regimes delimited
by the temperature $T^{*}$, 
one for $T_{f}<T<T^{*}$, and another one for $T>T^{*}$. 
It is argued that the conventional paramagnetic state corresponds
to $T>T^{*}$, whereas the temperature region $T_{f}<T<T^{*}$
may be characterized by a rather unusual dynamics, possibly 
including Griffiths singularities.   

\vskip \baselineskip
% 05.50.+q : Lattice theory and statistics (Ising, Potts, etc.) 
% 05.70.Fh : Phase transitions: general studies
% 05.70.Jk : Critical point phenomena
% 64.60.De : Statistical mechanics of model systems (Ising 
% model, Potts model, Monte Carlo techniques, etc.)
% 64.60.F- : Equilibrium properties near critical points, 
% critical exponents
% 75.10.Hk : Classical spin models
% 75.10.Nr : Spin glasses and other random models (theory)
% 75.50.Lk : Spin glasses and other random magnets (magnetic properties)

\end{abstract}

\maketitle
\section{Introduction}
% Abbreviations: SG  RF  SK  RS  RSB  QCP 
Nature is quantum in its essence,  although classical theories may be employed under certain conditions.
In statistical mechanics, the 
temperature range becomes crucial for the use of 
classical or quantum approaches. 
Typical examples appear in magnetism, where the use  
of classical models is justified when the temperature ranges are high enough, when 
compared to some reference temperature. 
In many magnetic systems the quantum effects become relevant, 
and should be taken into account, like those within the realm of
quantum magnetism~\cite{noltingbook}. 

In what concerns spin glasses (SGs), models based on Ising variables
have been able to describe fairly well, at least qualitatively, a wide 
variety of experimental behavior, even for sufficiently low 
temperatures~\cite{binderyoung,fischerhertzbook,dotsenkobook,%
nishimoribook}.
Some of these results have been obtained at mean-field level, based on 
the infinite-range-interaction Sherrington-Kirkpatrick (SK) model~\cite{SK75}, 
either by means of the replica-symmetric (RS), or replica-symmetry-breaking (RSB), 
solutions~\cite{Parisi80a}. 
Although this may seem paradoxical, due to the fact that Ising 
SG Hamiltonians are not formulated in terms of 
quantum operators, it is understood since their binary variables capture 
an essential ingredient of many physical systems, for which 
strong anisotropy fields are present, leading to two significant 
states associated with the spin operators. 
However, in some compounds, the quantum fluctuations controlled by a
given parameter (e.g., magnetic field, and/or doping) depress 
the transition temperature $T_f$, changing radically the physical 
properties of the system~\cite{sachdevbook}. 
In some cases, a field transverse to such spin operators
appears to be relevant, and so, the simpler Ising SG 
Hamiltonian should be replaced by a quantum type
of Hamiltonian. 

The Ising dipolar-coupled ferromagnet LiHoF$_4$ 
is a well known system in which quantum fluctuations become 
important by applying a transverse magnetic field $H_t$, 
which induces quantum tunnelling through the barrier separating 
the two degenerate ground states of Ho$^{3+}$ ions~\cite{i2}.
Moreover, disorder can be introduced, by replacing the magnetic Ho$^{3+}$ 
ions by nonmagnetic Y$^{3+}$ ones. Therefore,  the resulting 
LiHo$_x$Y$_{1-x}$F$_4$ compound is considered as an ideal 
ground for investigating the interplay between quantum fluctuations
and disorder in Ising spins systems~\cite{Wu91,Gingras11,Quiliam12}.

In these physical systems, coefficients of the expansion 
of the magnetization $m$, in powers of a small external longitudinal field 
$H_{l}$, are quantities of great 
interest~\cite{binderyoung,fischerhertzbook,Chalupa77},
\begin{equation}
m = \chi_{1}H_{l} - \chi_{3}H_{l}^{3} - \chi_{5}H_{l}^{5} - \cdots~,  
\label{mexpansion}
\end{equation} 
corresponding to the linear susceptibility, 
\begin{equation}
\chi_1= \left. \frac{ \partial m}{ \partial H_l} 
\right|_{H_l \rightarrow 0 }~,
\label{chi1}
\end{equation} 
and nonlinear susceptibilities, 
\begin{equation}
\chi_3= -\frac{1}{3!} \left. \frac{ \partial^3 m}{ \partial H_l^3} 
\right|_{H_l \rightarrow 0 }~; \quad   
\chi_5= -\frac{1}{5!} \left. \frac{ \partial^5 m}{ \partial H_l^5} 
\right|_{H_l \rightarrow 0 }~.  
\label{chi3}
\end{equation} 
Since measurements of $\chi_5$ (and higher-order susceptibilities) may
become a hard task, very frequently in the literature one refers to $\chi_3$
as the nonlinear susceptibility. Moreover, $\chi_3$ is directly related to the  
SG susceptibility~\cite{binderyoung,fischerhertzbook},  
\begin{equation}
\chi_3 = \beta^{2} \left( \chi_{\rm SG} - {2\over 3} \right)~, 
\label{chi3sgchi}
\end{equation} 
with the latter representing an important
theoretical tool, being defined as
\begin{equation}
\chi_{\rm SG} = {\beta \over N} \sum_{i,j} \left[
(\langle S_iS_j \rangle - \langle S_i \rangle \langle S_j \rangle)^2\right]_{\rm av}~, 
\label{sgchi}
\end{equation} 
where $\langle ..\rangle$ and $[..]_{\rm av}$
denote, respectively, thermal averages and an average over the disorder. 

In fact, the interplay between quantum fluctuations
and disorder stands for the physical origin of the intriguing 
behavior found in the magnetic susceptibility $\chi_3$ of 
LiHo$_x$Y$_{1-x}$F$_4$, which has been the object of 
a considerable experimental and theoretical debate. 
In the absence of $H_t$, the LiHo$_{0.167}$Y$_{0.833}$F$_4$ compound 
displays a sharp peak in $\chi_3$ 
at the temperature $T_f$, which resembles 
a conventional second-order SG phase transition~\cite{Wu93}.  
Surprisingly, the sharp peak of $\chi_3$ becomes increasingly 
rounded when the transverse field $H_t$ is applied and 
enhanced, so that the resulting smooth curve 
presents a maximum located at a temperature $T^*$, with $T^*>T_f$.
Such behavior was initially interpreted as a changing 
in the nature of the transition, from second order at $T_f$ 
to first order at $T^*$~\cite{Wu93,Cugliandolo01}.   
More recently, J\"onsson and collaborators~\cite{Jonsson07} 
investigated the behavior of $\chi_3$ 
for dopings $x=0.165$ and $0.0045$, obtaining the 
same roundings of the peak in both cases; these authors 
understood this behavior as an evidence of absence 
of a SG phase transition of any nature. 
In contrast,  Ancona-Torres and collaborators~\cite{Torres08} 
performed measurements for doping $x=0.167$, not only 
of $\chi_3$, but also of $\chi_5$, as well as of the ac susceptibility, 
reasserting $T^*$ as the SG critical temperature. 

On the theoretical side, the debate on this particular issue has also been 
intense (see, for instance, Refs.~\cite{Fernandez2010, Alonso2015}).
The suggestion that an effective longitudinal random field (RF) 
$h_i$ can be induced from the interplay of a transverse applied field 
$H_t$, with the off-diagonal terms of the dipolar interactions in 
LiHo$_x$Y$_{1-x}$F$_4$, represents a very interesting hint
to clarify these controversies concerning the meaning 
of $T^*$~\cite{Laflorencie06,Schechter1,Gingras06,Gingras08,Mydosh15}.
According to the droplet picture for SGs,  the rounded behavior of $\chi_3$ in 
the presence of the field-induced RF $h_i$ is interpreted as 
a suppression of the SG transition~\cite{Laflorencie06,Schechter1},
similar to what a uniform field does in that picture~\cite{Fisher86,Young04}.  
On the other hand, Tabei and collaborators~\cite{Gingras06} working 
within Parisi's mean-field theory~\cite{Parisi80a}, using an effective 
Hamiltonian defined in terms of the 
field-induced RF $h_i$ and a transverse field $\Gamma$
[where $\Gamma=\Gamma(H_t)$ represents some monotonically  
increasing function of $H_t$], 
reproduced quite well the experimental behavior of $\chi_3$, 
with an increasingly rounded peak at $T^*$ when 
$H_t$ is enhanced.

It should be remarked that the results described above are
based on a particular approach of the quantum SK model proposed by 
Kim and collaborators~\cite{Kim02}. In this approach, 
the SK model is analyzed in the presence of a transverse field $\Gamma$,  
within the static approximation. 
Mostly important, a region inside the SG phase was found  
where the RS approximation is stable. Actually, the RSB solution 
exists only for sufficiently low values of $\Gamma$, 
at temperatures lower than the SG transition temperature.
The main consequence of this scenario is that the sharp peak of 
$\chi_3$, which signals the SG phase transition temperature, 
does not coincide with the onset of RSB.
Nevertheless, this result is also highly controversial, since other works
indicate precisely the opposite, i.e., the RS approximation
is unstable  throughout the whole SG phase (see, for instance,
Refs.~\cite{Goldschmidt90,Huse93}) except, possibly, at the zero-temperature 
Quantum Critical Point (QCP)~\cite{Sachdev95}. 
Consequently, when $\Gamma$ enhances, the RSB transition temperature 
$T_f$ decreases, so that, for finite temperatures,  
the critical behavior appears in $\chi_3$ as 
\begin{equation}
\chi_3\propto [(\Gamma-\Gamma_f(T))/\Gamma_f(T)]^{-\delta^{\prime}}~,  
\label{x3critic}
\end{equation} 
where $\Gamma_f(T)$ denotes the critical value of $\Gamma$ for a
given temperature, from which its corresponding value at the 
zero-temperature QCP is 
obtained as $\lim_{T \rightarrow 0}\Gamma_f(T)=\Gamma_c^0$. 

In the classical case, it is well known that $\chi_{\rm SG}$ is 
inversely proportional to the Almeida-Thouless eigenvalue 
$\lambda_{\rm AT}$, the so-called replicon~\cite{binderyoung,Almeida78}.  
Therefore, the diverging behavior of $\chi_3$ at the SG transition, 
\begin{equation}
\chi_3\propto [(T-T_f)/T_f]^{-\gamma}~,  
\label{x3critic1}
\end{equation}
is a direct consequence of $\lambda_{\rm AT}=0$ at $T_f$, 
occurring together with the onset of RSB. Similarly, in the quantum case,  
one expects that the divergence of $\chi_3$ at $\Gamma_f(T)$ 
should coincide with the onset of RSB.  

Indeed, the presence of a RF can produce deep 
changes in the scenario described previously.
For instance, in the classical SK model~\cite{SK75}, 
the RF induces the RS order parameter $q$, which 
becomes finite at any temperature~\cite{Pytte77,Krzakala10}. 
As a consequence, $q$ versus temperature presents 
a smooth behavior, being no more appropriate 
for identifying a SG transition in the SK model.
Nevertheless, such a transition may still be 
related with the onset of RSB, signaled by 
$\lambda_{\rm AT}=0$~\cite{SNA}. In spite of this, 
the derivative of $q$ with respect to the 
temperature increases as one approaches $T_{f}$ 
from above; such an increase is the ultimate 
responsible for the rounded maximum in $\chi_3$ 
at the temperature $T^*$, which does not coincide 
with the SG transition temperature $T_f$ ($T^* > T_f$).  
In fact, the maximum value of $\chi_3$ at $T^*$ 
reflects the effects of the RF inside the 
paramagnetic phase, instead of the non-trivial 
ergodicity breaking of the SG phase transition~\cite{CAVJ16}. 
Therefore, one can raise the question of whether 
such scenario for $\chi_3$, found in the classical SK model,  
is robust in the corresponding quantum model, 
when the transverse field is considered, i.e., 
$\Gamma \neq 0$, and no longer independent from the 
RF, as proposed by Tabei and collaborators~\cite{Gingras06}.

The purpose of the present work is to study the susceptibility $\chi_3$,   
using the so-called fermionic Ising SG model 
in the presence of  
a longitudinal RF $h_i$ and a transverse field $\Gamma$. 
In this model, the spin operators are written in terms of 
fermionic occupation and destruction 
operators~\cite{Theumann84,Opperman93}, whereas the 
spin-spin couplings $\{ J_{ij} \}$ and random fields $\{ h_i \}$ follow 
Gaussian distributions. 
The grand-canonical potential is obtained 
in the functional integral formalism, and the disorder is treated using the
replica method; moreover, the SG order parameters are obtained in the 
static approximation~\cite{Bray80,Thihumalai89} and investigated within the 
one-step RSB scheme~\cite{Parisi80a}. 
It should be remarked that the fermionic Ising SG model 
is defined on the Fock space, where there are four possible states 
per site: one state with no fermions, two states with a single 
fermion, and one state with two fermions, leading to  
two nonmagnetic states. In particular, one can consider two 
cases: the $4S$ model that allows the four possible states per site 
and the $2S$ model, which restricts the spin operators to act 
on a space where the nonmagnetic states are forbidden.
In the present work we will consider the later model, by imposing a 
restriction to remove the contribution of these nonmagnetic states, 
i.e., taking into account only the sites occupied by one fermion 
in the partition-function trace~\cite{wener86,zimmer06}.

In order to deal appropriately with the experimental behavior 
of $\chi_3$ in the LiHo$_x$Y$_{1-x}$F$_4$ compound, we focus 
our calculations on the 2S model by proposing a relationship between 
$\Delta$ (the width of the distribution of random fields $h_i$) 
and $\Gamma$, following the approach introduced
by Tabei and collaborators~\cite{Gingras06}. 
The main characteristic observed experimentally in $\chi_3$ 
concerns the peak for small $H_t$ (classical limit) being replaced 
by a rounded maximum which becomes increasingly rounded 
for large $H_t$ (quantum limit). Besides the progressive smearing of $\chi_3$, 
the amplitude of its maximum also decreases as $H_t$ increases. 
Therefore, the effects of the RF triggered by $H_t$, 
as suggested by Tabei and collaborators \cite{Gingras06}, 
should provoke simultaneously both effects, i.e., the smearing of the peak 
and the decrease of the maximum amplitude value of $\chi_3$. 
For the present fermionic Ising SG model, we have tested 
a relationship involving $\Delta$ and $\Gamma$, particularly  
in the power-like form, 
$\Delta/J \propto (\Gamma/J)^{B'}$, where $J$ represents the width of
the Gaussian distribution for the couplings $\{ J_{ij} \}$. 
Considering the interval for the exponent, $1.8<B'<2.5$,
we have been able to obtain $\chi_3$ as a function of temperature 
and $\Gamma$ resembling qualitatively
the experimental behavior for $\chi_3$ described above. 
As already mentioned, the
transverse field $\Gamma$ used in the effective model to describe  
LiHo$_x$Y$_{1-x}$F$_4$ is expected to be related to  
the experimental applied field $H_t$~\cite{Gingras11,Quiliam12};  
in fact, at least for low $H_t$,
$\Gamma \propto H_t^{2}$ (see, e.g., Ref.~\cite{Wu1991}). 

The paper is structured as follows: in Section II we define the model and
find its grand-canonical potential within the one-step RSB scheme;  
in Section III we present a detailed discussion of the order 
parameters, the susceptibility $\chi_3$, and some 
phase diagrams. Finally, the last section is reserved to conclusions.

\section{Model}

The model is defined by the Hamiltonian
\begin{equation}
H = - \sum_{(i,j)} J_{ij} \hat S_{i}^{z} \hat S_{j}^{z} 
- \sum_{i=1}^{N} h_{i} \hat S_{i}^{z}- 2\Gamma \sum_{i=1}^{N}\hat S_{i}^{x}, 
\label{eq1}
\end{equation}
where the summation $\sum_{(i,j)}$ applies to all distinct pairs of 
spin operators, 
whereas the couplings $\{ J_{ij} \}$ and magnetic fields $\{ h_i \}$ 
are quenched random variables, following independent 
Gaussian distributions, 
\begin{equation}
P(J_{ij})=\left[\frac{N}{32\pi J^{2}}\right]^{1/2} 
\exp\left[-\frac{N}{32J^{2}}\left(J_{ij} -J_0/N \right)^{2}\right]~,
\label{eq2a}
\end{equation}
and
\begin{equation}
P(h_i)=\left[\frac{1}{32\pi \Delta^{2}}\right]^{1/2} 
\exp\left[-\frac{1}{32\Delta^{2}} h_i^{2} \right]~. 
\label{eq2}
\end{equation}
In order to obtain susceptibilities [cf. Eqs.~(\ref{chi1}) and~(\ref{chi3})], 
one introduces a longitudinal uniform 
field $H_l$, by adding an extra term $- \sum_{i=1}^{N} H_{l} \hat S_{i}^{z}$ 
in the Hamiltonian above.
Moreover, the spin operators in Eq.~(\ref{eq1}) are defined as
\begin{equation}
\hat S_{i}^{z}=\frac{1}{2}[\hat{n}_{i\uparrow}-\hat{n}_{i\downarrow}]~; 
\qquad 
\hat S_{i}^{x}=\frac{1}{2}[\hat{c}_{i\uparrow}^{\dagger}\hat{c}_{i\downarrow}
+\hat{c}_{i\downarrow}^{\dagger}\hat{c}_{i\uparrow}]~, 
\label{ope}
\end{equation}
where 
$\hat{n}_{i\uparrow}=\hat{c}_{i\uparrow}^{\dagger}\hat{c}_{i\uparrow}$ and 
$\hat{n}_{i\downarrow}=\hat{c}_{i\downarrow}^{\dagger}\hat{c}_{i\downarrow}$, 
with $\hat{c}_{i\uparrow}^{\dagger}$ denoting a creation operator for 
a fermion with spin up at site $i$, $\hat{c}_{i\downarrow}$ an annihilation operator 
for a fermion with spin down at site $i$, and so on. 
In this fermionic problem, the partition function is expressed 
by using the Lagrangian path integral formalism in terms of 
anticommuting Grassmann fields ($\phi$ and $\phi^*$)~\cite{Opperman93}.
The restriction in the 2S-model is imposed by means of a 
Kronecker delta function, in such a way to take into account 
only those sites occupied by one fermion
($n_{i\uparrow}+n_{i\downarrow}=1$)
in the partition-function~\cite{wener86,zimmer06}.
Therefore, adopting an integral representation for this delta function, 
one can express the partition function for both $2S$ and 
$4S$ models in the following form, 
\begin{equation}
Z\{y\}=\mbox{e}^{\frac{s-2}{2}N\beta\mu}\int D(\phi^* \phi)\prod_{j}^{}\frac{1}{2\pi}\int_{0}^{2\pi}dx_j\mbox{e}^{-y_j}\mbox{e}^{A\{y\}}~, 
\end{equation}
where
\begin{equation}
\begin{split}
A\{y\}&=\int_{0}^{\beta}d\tau
\left\{
\sum_{j,\sigma}\phi_{j\sigma}^{*}(\tau)\left[
 \frac{\partial}{\partial \tau} +\frac{y_j}{\beta}\right]\phi_{j\sigma}(\tau) 
\right. \\& \left.   
- H\left(\phi_{j\sigma}^{*}
 (\tau),\phi_{j\sigma}(\tau)\right) \right\}
\label{action}~. 
\end{split}
\end{equation}
In the equations above, $\beta=1/T$ ($T$ being the temperature), 
$y_{j}=ix_j$ for the $2S$-model, or $y_{j}=\beta\mu$ for the $4S$-model, 
$s=2,4$ denotes the number of states per site allowed in each model, 
respectively, and $\mu$ is the chemical potential. 
Moreover,  
$\phi_{j\sigma}$ and $\phi_{j\sigma}^{*}$ represent 
Grassmann fields at site $j$ and spin state $\sigma$, whereas
$H\left(\phi_{j\sigma}^{*}(\tau),\phi_{j\sigma}(\tau)\right)$ stands
for an effective Hamiltonian at a given value of the integration
variable $\tau$. 

Now, we use the replica method, so that standard procedures
lead to the grand-canonical potential per particle~\cite{Fisher86}, 
\begin{equation}
\beta \Omega=-\frac{1}{N}\langle \langle\ln Z\{y\}\rangle\rangle_{J,h}=-\frac{1}{N}\lim_{n\longrightarrow 0}
\frac{\langle\langle Z\{y\}^n \rangle\rangle_{J,h} - 1}{n}~, 
\label{omega}
\end{equation}
where $\langle\langle .. \rangle\rangle_{J,h}$ denote averages over 
the quenched random variables.
The replicated partition function  
$\langle\langle Z\{y\}^n \rangle \rangle_{J,h}$ becomes
\begin{equation}
\begin{split}
\langle\langle Z\{y\}^n\rangle\rangle_{J,h}&=\mbox{e}^{\frac{s-2}{2}N\beta\mu}
{\cal N}\int_{-\infty}^{\infty} \prod_{(\alpha,\gamma)} dq_{\alpha\gamma} 
\int_{-\infty}^{\infty} \prod_{\alpha=1}^{n} dq_{\alpha\alpha} 
\\& \times \int_{-\infty}^{\infty} \prod_{\alpha=1}^{n} dm_{\alpha} 
\exp\left[N {\beta}\Omega_{n}(q_{\alpha\gamma}, q_{\alpha\alpha},m_{\alpha}) \right]
\end{split}
\label{qqm}
\end{equation}
where $\alpha$ ($\alpha=1,2,\cdots,n$) stands for a replica index, 
$(\alpha,\gamma)$ denotes distinct pairs of replicas, and    
${\cal N}=(\beta J\sqrt{N/2\pi})^{n(n+1)/2}$.   
Assuming the static approximation~\cite{Bray80,Thihumalai89}, one obtains
\begin{equation}
\begin{split}
{\beta}\Omega_{n}(q_{\alpha\gamma}, q_{\alpha\alpha},m_{\alpha})
=-{\beta^2J^2}\sum_{(\alpha,\gamma)}q_{\alpha\gamma}^{2}
\\-\frac{\beta^2J^2}{2}\sum_{\alpha}q_{\alpha\alpha}^{2} - \frac{\beta J_0}{2} \sum_\alpha m_{\alpha}^{2} + \ln\Lambda\{y\}~, 
\end{split}
\label{stacionario}
\end{equation}
and the Fourier representation may be used to express
\begin{equation}
\Lambda\{y\}=\prod_{\alpha}\frac{1}{2\pi}\int_{0}^{2\pi}dx_{\alpha}\mbox{e}^{-y_{\alpha}}
\int D[\phi_{\alpha}^{*},\phi_{\alpha}]\exp[H_{\rm eff}]~. 
\label{efetivo}
\end{equation}
Above, one has an ``effective Hamiltonian'' in replica space, 
\begin{equation}
\begin{split}
H_{\rm eff}&=\sum_{\alpha} A_{0\Gamma}^{\alpha} + 4 \left[ \frac{\beta^{2}\Delta^{2}}{2}\sum_{\alpha,\gamma}S_{\alpha}^{z}S_{\gamma}^{z} 
+ \frac{\beta J_0}{2} \sum_\alpha m_{\alpha} S_{\alpha}^{z} \right. \\& + \left. \beta^{2}J^{2}\left( 
\sum_{\alpha}	q_{\alpha\alpha}S_{\alpha}^{z}S_{\alpha}^{z}
+2\sum_{(\alpha,\gamma)}q_{\alpha\gamma}S_{\alpha}^{z}S_{\gamma}^{z} \right)  \right]~, 
\end{split}
\label{heff}
\end{equation}
with 
\begin{equation} 
\begin{split}
A_{0\Gamma}^{\alpha}&=\sum_{\omega}\underline{\varphi}_{\alpha}^{\dagger}(\omega)(i\omega+y_\alpha+
\beta\Gamma \underline{\sigma}^{x})
\underline{\varphi}_{\alpha}(\omega),\\& ~~S_{\alpha}^{z}=\frac{1}{2}\sum_{\omega}
\underline{\varphi}_\alpha(\omega)\underline{\sigma}^{z}\underline{\varphi}_\alpha(\omega),
\end{split}
\end{equation}
where the Matsubara's frequencies are  $\omega=\pm \pi,\pm 3\pi,\cdots $, 
$\underline{\sigma}^x$ and $\underline {\sigma}^z$ denote the Pauli matrices, and 
$\underline{\varphi}_{\alpha}^{\dagger}(\omega)
=\left(\phi_{\uparrow\alpha}^{*}(\omega)
~~\phi_{\downarrow\alpha}^{*}(\omega)\right)$.

Moreover, the functional integrals over $q_{\alpha\gamma}$, 
$q_{\alpha\alpha}$ and $m_{\alpha}$ in Eq.~(\ref{qqm}) have been
evaluated through the steepest-descent method, 
yielding 
\begin{equation}
m_{\alpha}=\langle S_{\alpha} \rangle~; \quad 
q_{\alpha\gamma}=\langle S_{\alpha}^{z}S_{\gamma}^{z}\rangle~; 
\quad 
q_{\alpha\alpha}=\langle (S_{\alpha}^{z})^2\rangle~,
\end{equation}
with $\langle ..\rangle$ representing a thermal average over the effective 
Hamiltonian of Eq.~(\ref{heff}).  

Herein, the problem will be analyzed within one-step RSB Parisi's 
scheme~\cite{Parisi80a}, in which $q_{\alpha\alpha}=p$, and 
the replica matrix elements are parametrized as 
\begin{equation}
q_{\alpha\gamma}=\begin{cases}
q_{1} & \mbox{if}\ I(\alpha/a)=I(\gamma/a)\\
q_{0} & \mbox{if}\ I(\alpha/a)\ne I(\gamma/a)
\end{cases}\label{rsb}
\end{equation}
where $I(x)$ gives the smallest integer greater than, or
equal to $x$. 

The parametrization given by Eq.~(\ref{rsb}) allows to perform the 
sums over replica indexes and then, the quadratic terms in Eq.~(\ref{heff}) 
can be linearized through the introduction of new auxiliary fields. 
From this point, the integrals over the Grassmann variables in 
Eq.~(\ref{efetivo}) can be performed and the sum over 
Matsubara's frequencies 
can be obtained, like in Ref.~\cite{zimmer06}. Therefore, 
the resulting grand-canonical potential is obtained 
from Eq.~(\ref{omega}), 
\begin{equation}
\begin{split}
\beta \Omega&=\frac{(\beta J)^{2}}{2}[(x-1)q_{1}^{2}-x q_{0}^{2}+p^{2}] + \frac{\beta J_0}{2}m^2 -
\ln 2\\ & -\frac{(s-2)}{2}\beta\mu -\frac{1}{x}\int Dz\ln \left\lbrace \int Dv [K
(z,v)
]^{x} \right\rbrace~,  
\label{grandpot}
\end{split}
\end{equation}
where 
\begin{equation}
\begin{split}
K(z,v)&=\frac{(s-2)}{2}\cosh(\beta\mu)+\int D\xi 
\cosh[\sqrt{\Xi(z,v,\xi)}], \\
\Xi(z,v,\xi)&=[\beta h(z,v,\xi)]^{2}+(\beta\Gamma)^{2}, \\
h(z,v,\xi)&=\beta J[ \sqrt{ 2 q_0 + (\Delta/J)^2} \, z + \sqrt{2 (q_1-q_0)} \, v
\\& + \sqrt{2 (p-q_1)} \, \xi],
\end{split}
\label{e30sgs}
\end{equation}
and $Dx \equiv dx \ $e$^{-x^2/2}/\sqrt{2\pi}$ ($x=z,~v$ or $\xi$).

The parameters $q_0$, $q_1$, $x$, $p$, and $m$ are 
obtained through extremization of the grand-canonical potential in  
Eq.~(\ref{grandpot}), and results for 2S and 4S models are obtained 
by considering $s=2$ and $s=4$, respectively. Moreover, the 
RS solution is recovered for $q_0=q_1= q$, and $x=0$. 
In this way, the linear susceptibility of Eq.~(\ref{chi1}) becomes
$\chi_1 = \beta[p -q_1+ x(q_1-q_0)]$~\cite{Parisi80a}.  

As usual, the RSB parameters, the magnetization 
$m$, and the quadrupolar parameter $p$, form a set 
of coupled equations, to be solved simultaneously. 
Particularly, the parameter $p$ is quite dependent on $\Gamma$, 
and in fact, for the 2S model, 
$p\rightarrow 1$ only as $\Gamma\rightarrow 0$; it should be mentioned
that $p$ plays an important role in the 
nonlinear susceptibility $\chi_3$.
This aspect represents a crucial difference of
the present investigation with respect to previous one, by Kim and 
collaborators [cf. Ref.~\cite{Kim02}], where the parameters 
$q_1$, $q_0$, and $x$ (or even $q$ in the RS solution) do 
not depend on $p$. 
In the present work, for the above one-step RSB solution, $\chi_3$ 
will be obtained 
by numerical derivatives; for the RS solution, an analytical form for  
$\chi_3$ is presented in Appendix A. 
As mentioned before, in order to deal with the LiHo$_x$Y$_{1-x}$F$_4$ 
compound, we will restrict ourselves to the 2S model, 
considering $J_0=0$. 

%%%%%%%%%%%%%%%%%%%%%%%%%
\begin{figure}[htb]
\includegraphics[width=6cm,angle=-90]{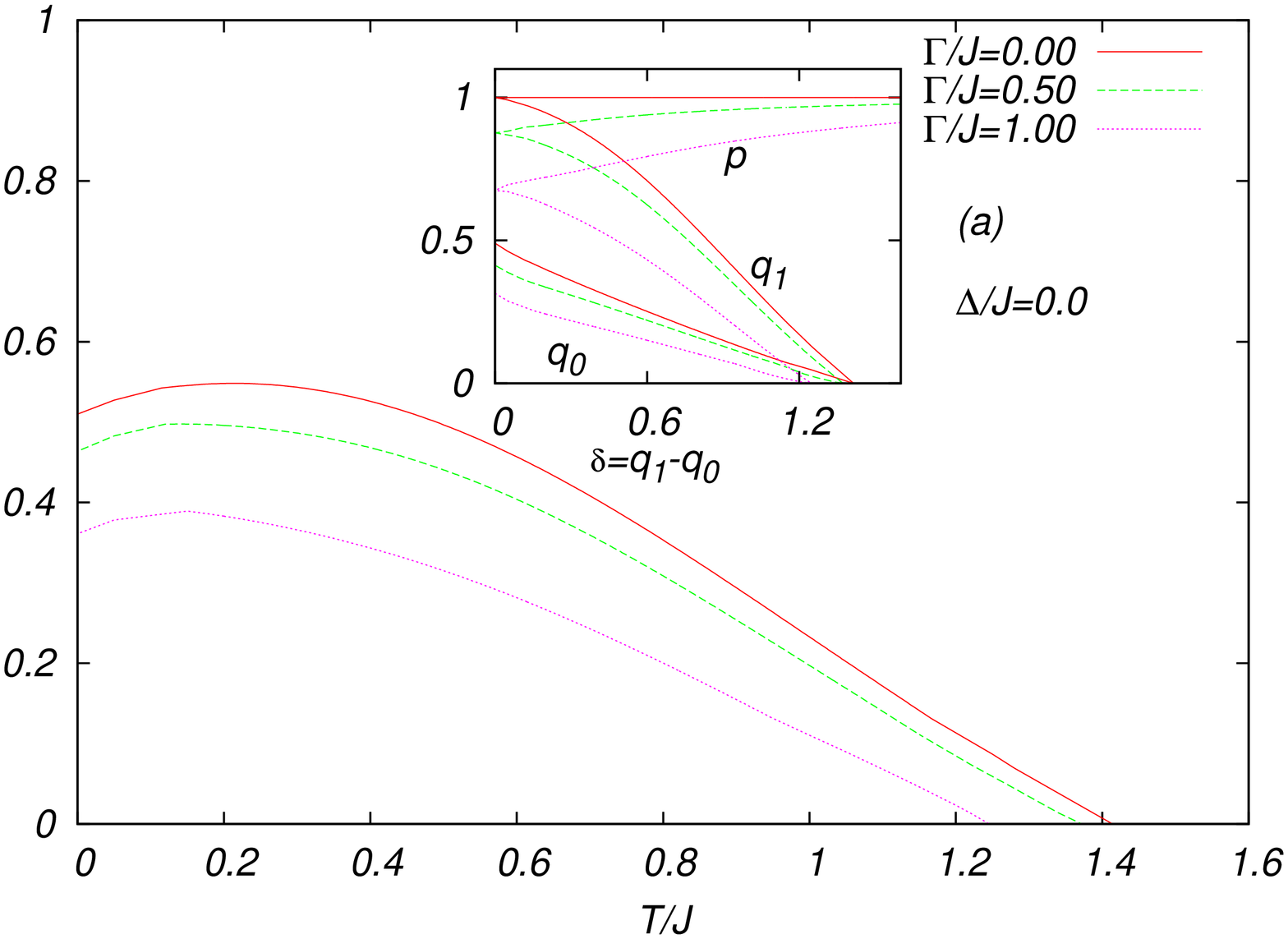} 
\includegraphics[width=6cm,angle=-90]{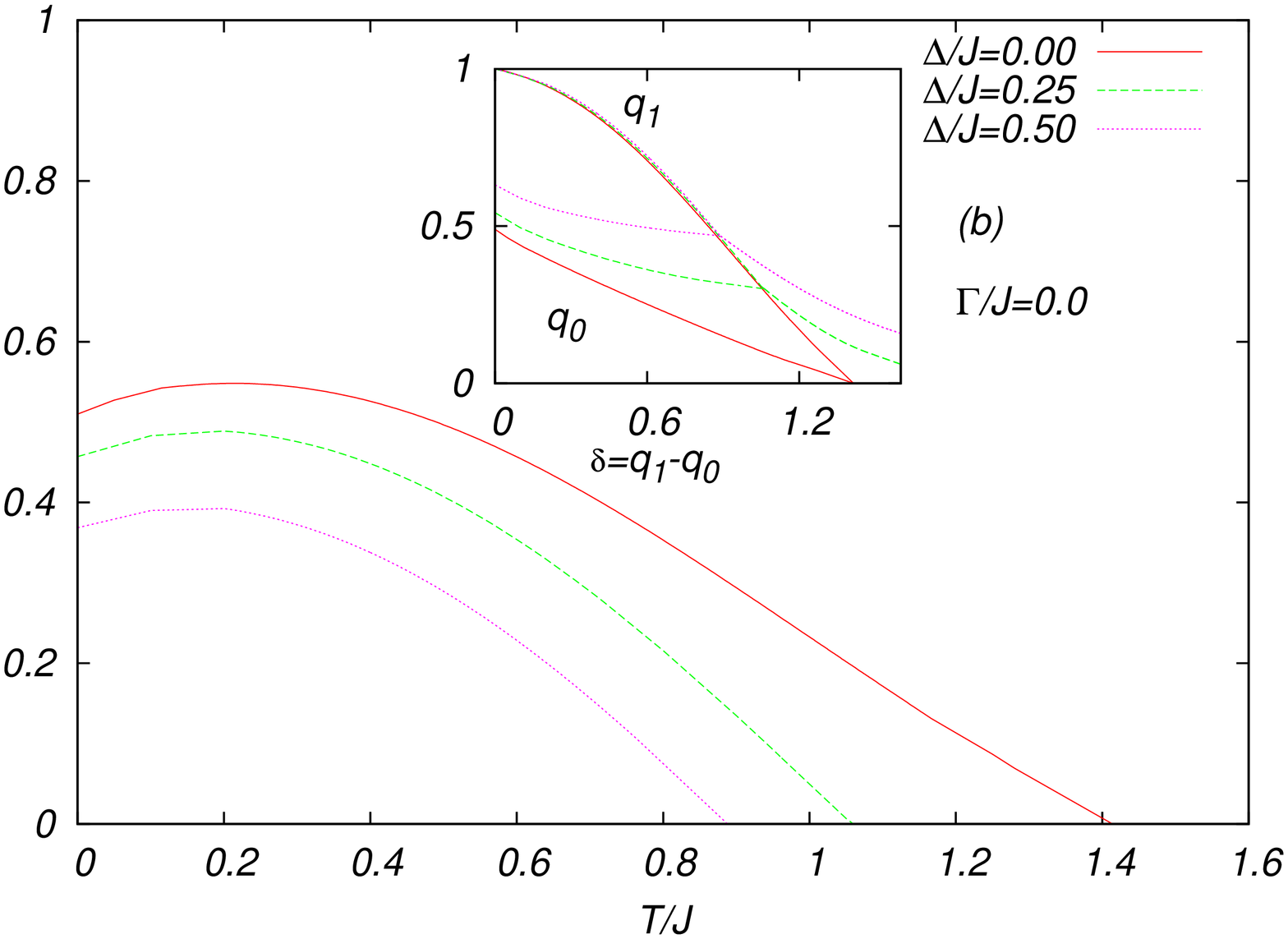} 
\caption{ 
The one-step RSB parameter $\delta\equiv q_1-q_0$ is presented 
versus the dimensionless temperature $T/J$, for $\Delta=0$ and typical 
values of $\Gamma/J$ [panel (a)], as well as for $\Gamma=0$ and
typical values $\Delta/J$ [panel (b)]. 
The parameters $q_1$, $q_0$, and $p$ are also exhibited versus 
temperature in the respective insets; one notices that the  
quadrupolar parameter $p$
becomes relevant only in the cases $\Gamma>0$, for which it 
decreases by lowering the temperature.   
Due to the usual numerical difficulties, the 
low-temperature results [typically $(T/J)<0.05$)] correspond to 
smooth extrapolations from higher-temperature data. 
}
\label{fig1}
\end{figure}
%%%%%%%%%%%%%%%%%%%%%%%%%

\section{Results}

Hence, considering the 2S  model, in this section we analyze the behavior 
of the nonlinear susceptibility $\chi_3$, either by varying the temperature 
(for fixed typical values of 
$\Delta/J$ and $\Gamma/J$), or by considering joint variations in some 
of these parameters. Since $\chi_3$ is directly related with the order 
parameters that appear in 
the thermodynamic potential of Eq.~(\ref{grandpot}), we first discuss 
the SG order parameters $q_1$ and $q_0$, as well as the quadrupolar 
parameter $p$. Moreover, the onset of RSB is signalled by 
$\delta=q_1-q_0>0$, which locates the freezing temperature $T_f$; it
should be mentioned that for $\Gamma=0$ and $\Delta=0$, one has 
that $T_f=\sqrt{2}J$.

%%%%%%%%%%%%%%%%%%%%%%%%%
\begin{figure*}[htb]
\includegraphics[width=6cm,angle=-90]{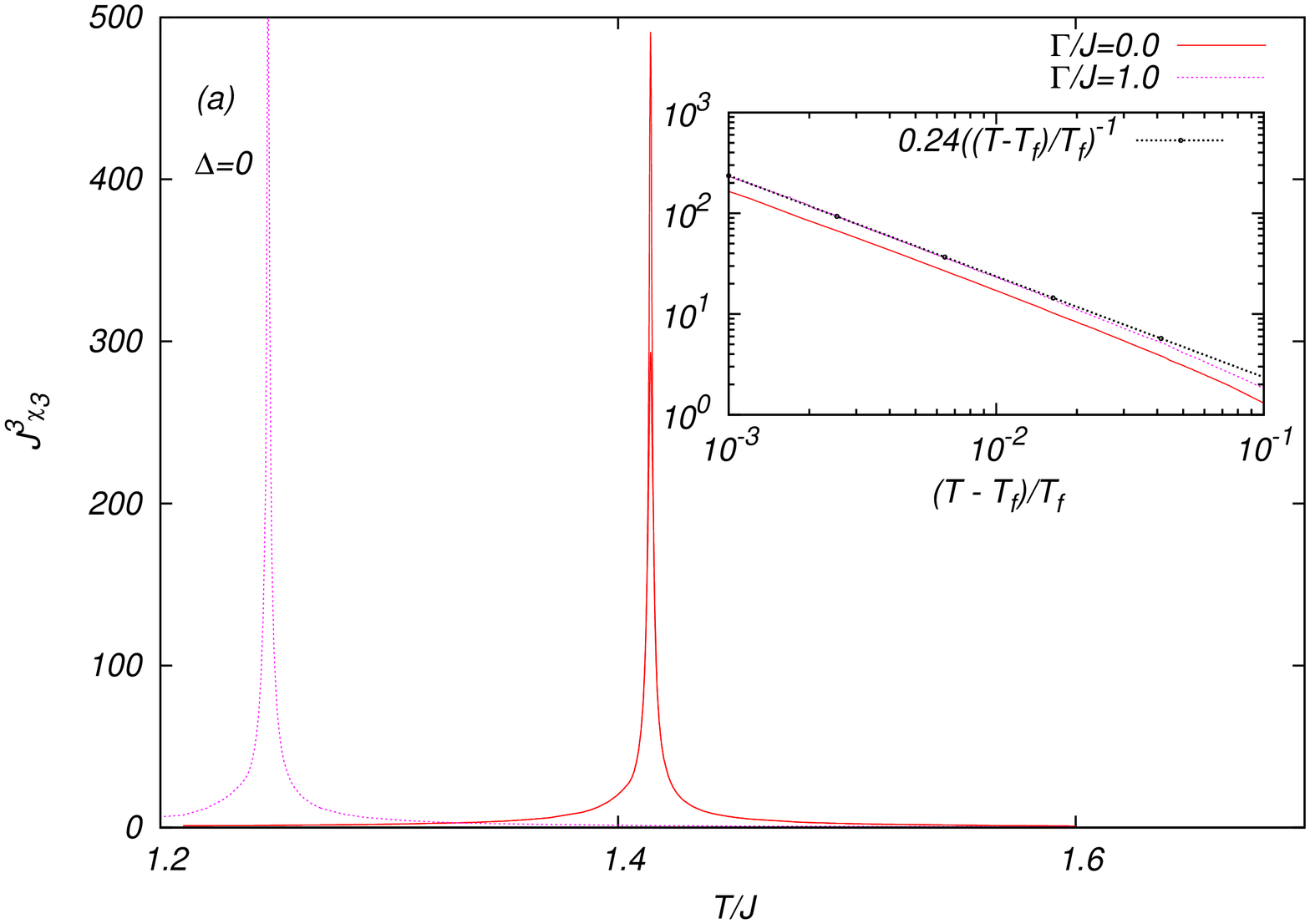}
\includegraphics[width=6cm,angle=-90]{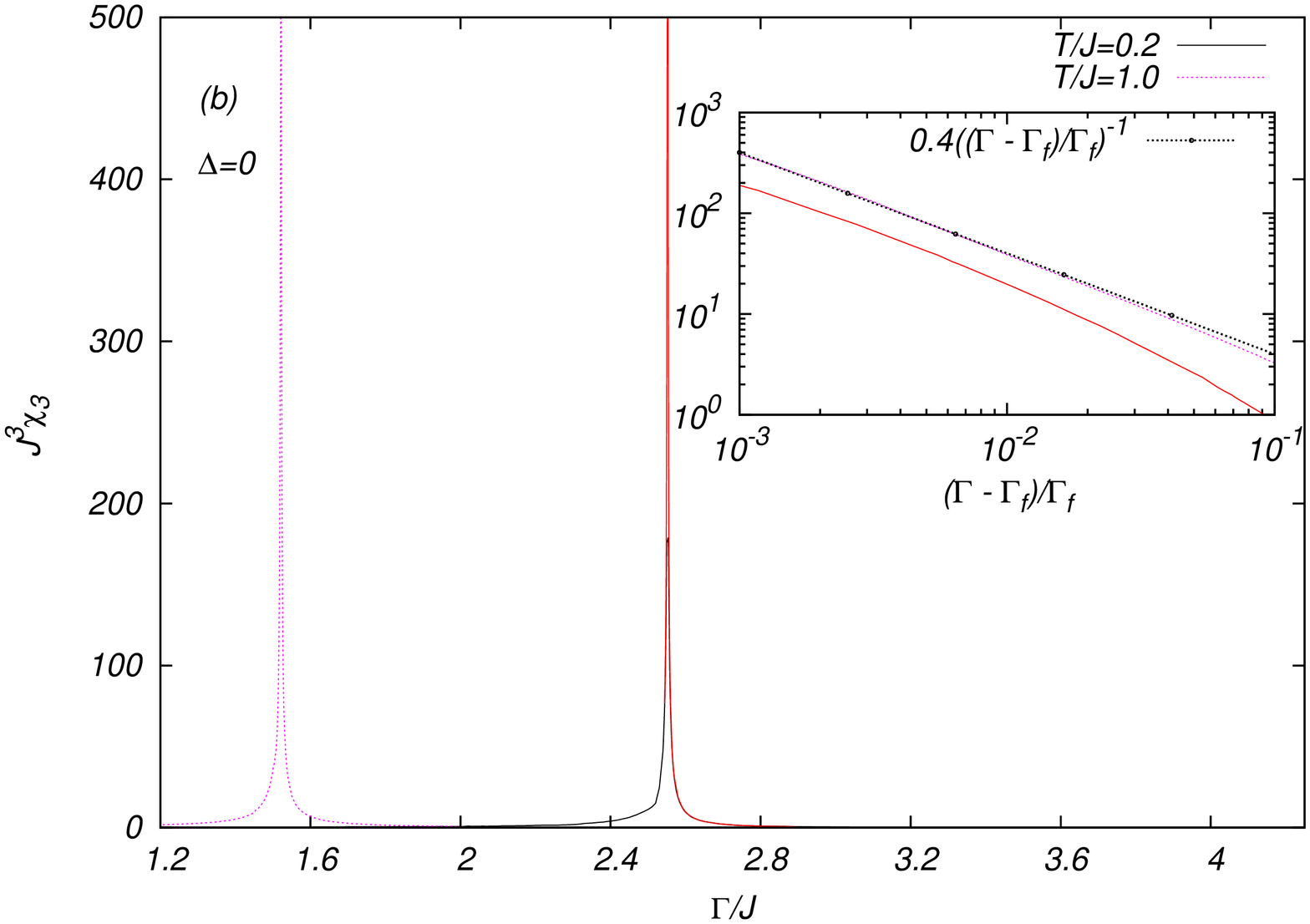} 
\caption{ 
Plots of the dimensionless nonlinear susceptibility 
[computed from Eq.~(\ref{chi3})]
are exhibited for $\Delta=0$: 
(a) $ J^{3}\chi_{3}$ versus $T/J$ for two different values of $\Gamma/J$;
(b) $ J^{3}\chi_{3}$ versus $\Gamma/J$ for two different temperatures.
In all cases one notices sharp divergences of $\chi_{3}$, signaling  
evident phase transitions. The corresponding critical exponents
are estimated through log-log plots (shown in the respective insets), where 
 in each case, the fitting proposal is represented by a dashed-dotted line (see text). 
}
\label{fig2}
\end{figure*}
%%%%%%%%%%%%%%%%%%%%%%%%% 

%%%%%%%%%%%%%%%%%%%%%%%%%
\begin{figure*}[htb]
\includegraphics[width=6cm,angle=-90]{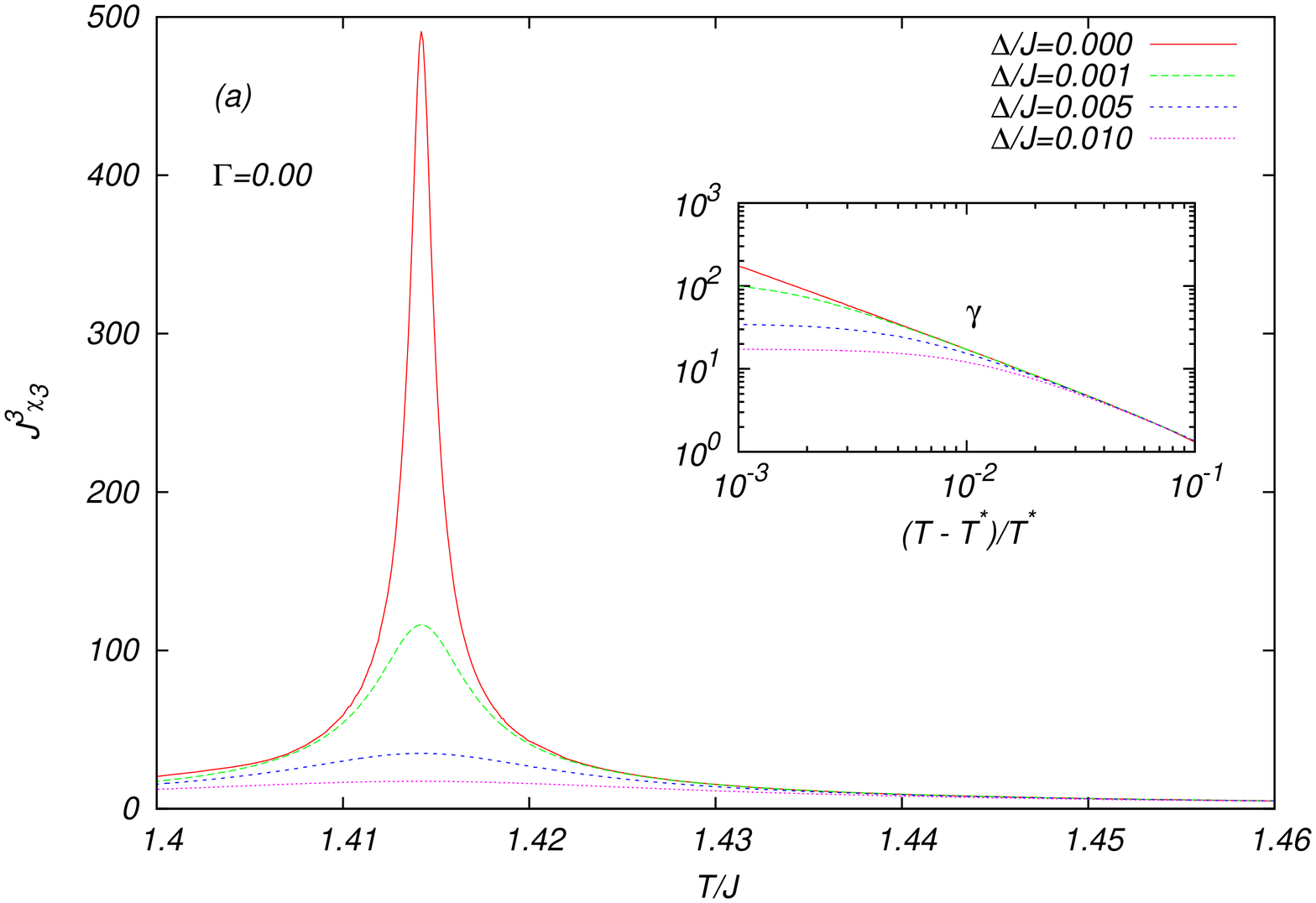}
\includegraphics[width=6cm,angle=-90]{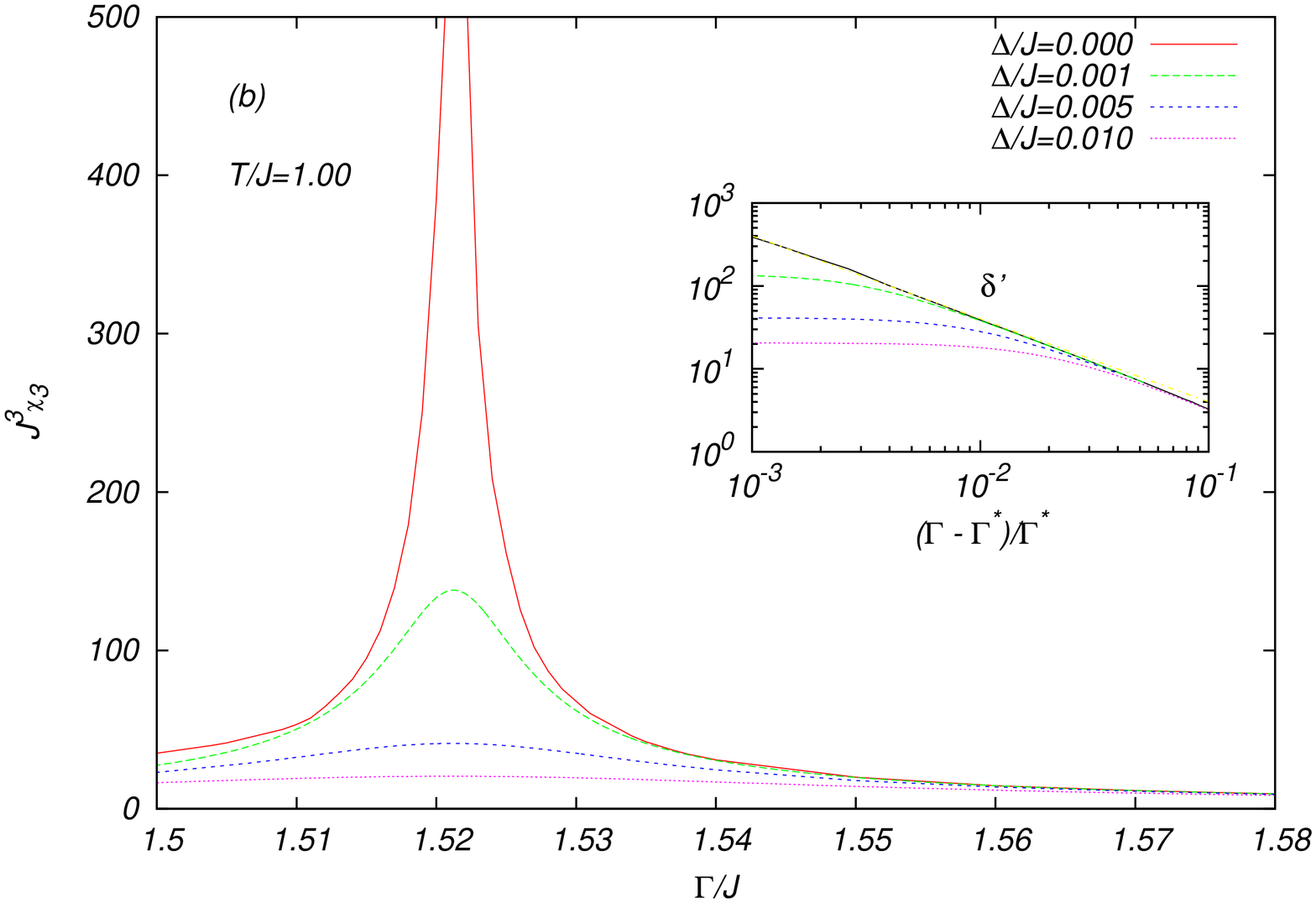} 
\caption{ 
The behavior of the dimensionless nonlinear susceptibility 
(in two typical cases exhibited in Fig.~\ref{fig2})
is presented for increasing values of $\Delta/J$: 
(a) $ J^{3}\chi_{3}$ versus $T/J$, for $(\Gamma/J)=0.0$; 
(b) $ J^{3}\chi_{3}$ versus $\Gamma/J$, for $(T/J)=1.0$.
In each case one notices a rounded peak for $(\Delta/J)>0$, with its maximum value
located at a temperature $T^{*}$ [panel (a)], or at a transverse field $\Gamma^{*}$ [panel (b)], such
that its height decreases for increasing values of $\Delta/J$. The log-log plots in the respective insets 
show that the divergences of Eq.~(\ref{x3critic1}), leading to the exponent $\gamma$ [inset of (a)], 
or in Eq.~(\ref{x3critic}), leading to the exponent  $\delta^{\prime}$ [inset of (b)],
are fulfilled only for $(\Delta/J)=0$.   
}
\label{fig3}
\end{figure*}
%%%%%%%%%%%%%%%%%%%%%%%%%

In Fig.~\ref{fig1} we exhibit the one-step RSB parameter $\delta\equiv q_1-q_0$ 
versus the dimensionless temperature $T/J$, for typical choices of 
$\Gamma/J$ and $\Delta/J$. The corresponding parameters 
$q_1$, $q_0$, and $p$ are also presented versus $T/J$ in the 
respective insets. From Fig.~\ref{fig1}(a) one notices that the freezing
temperature gets lowered for increasing values of the transverse
field $\Gamma$, up to the zero-temperature QCP located 
at $\Gamma_c^{0}=2\sqrt{2} J$ 
($\Delta=0$)~\cite{zimmer06}; a similar effect is verified in   
Fig.~\ref{fig1}(b) by increasing the width of the distribution of
random fields $\Delta$ ($\Gamma=0$).  
In the Hamiltonian of Eq.~(\ref{eq1}) one sees that the 
limit $\Gamma=0$ corresponds to a simple, diagonalizable,
quantum Ising SG model, where only the spin components 
$\hat S_{i}^{z}$ are present, leading to a trivial quadrupolar parameter
$p=1$ (for all temperatures), as shown in the inset of Fig.~\ref{fig1}(a).  
This particular case, for which the SG parameters are exhibited in
Fig.~\ref{fig1}(b), yields results qualitatively similar to those
found in the previous study of the classical SK model in the 
presence of a Gaussian random field, carried in Ref.~\cite{CAVJ16}.  
One notices that for $(\Delta/J)>0$ the RS order parameter 
$q=q_0=q_1$ is induced [cf. the inset of Fig.~\ref{fig1}(b)], 
presenting a smooth behavior versus temperature;
consequently, the freezing temperature $T_f$ can only be found by 
means of the RSB scheme, with the SG transition coinciding 
with the onset of the parameter $\delta$.   
However, for $\Gamma>0$, the spin components 
$\hat S_{i}^{x}$ become important, so that one expects 
a nontrivial behavior for the quadrupolar parameter $p$; indeed, 
$p$ should decrease for increasing values of $\Gamma$ (at a fixed
temperature), whereas for a fixed $\Gamma$, it decreases by lowering 
the temperature, as shown in the inset of Fig.~\ref{fig1}(a).

%%%%%%%%%%%%%%%%%%%%%%%%%
\begin{figure*}[htb]
\includegraphics[width=6.1cm,angle=-90]{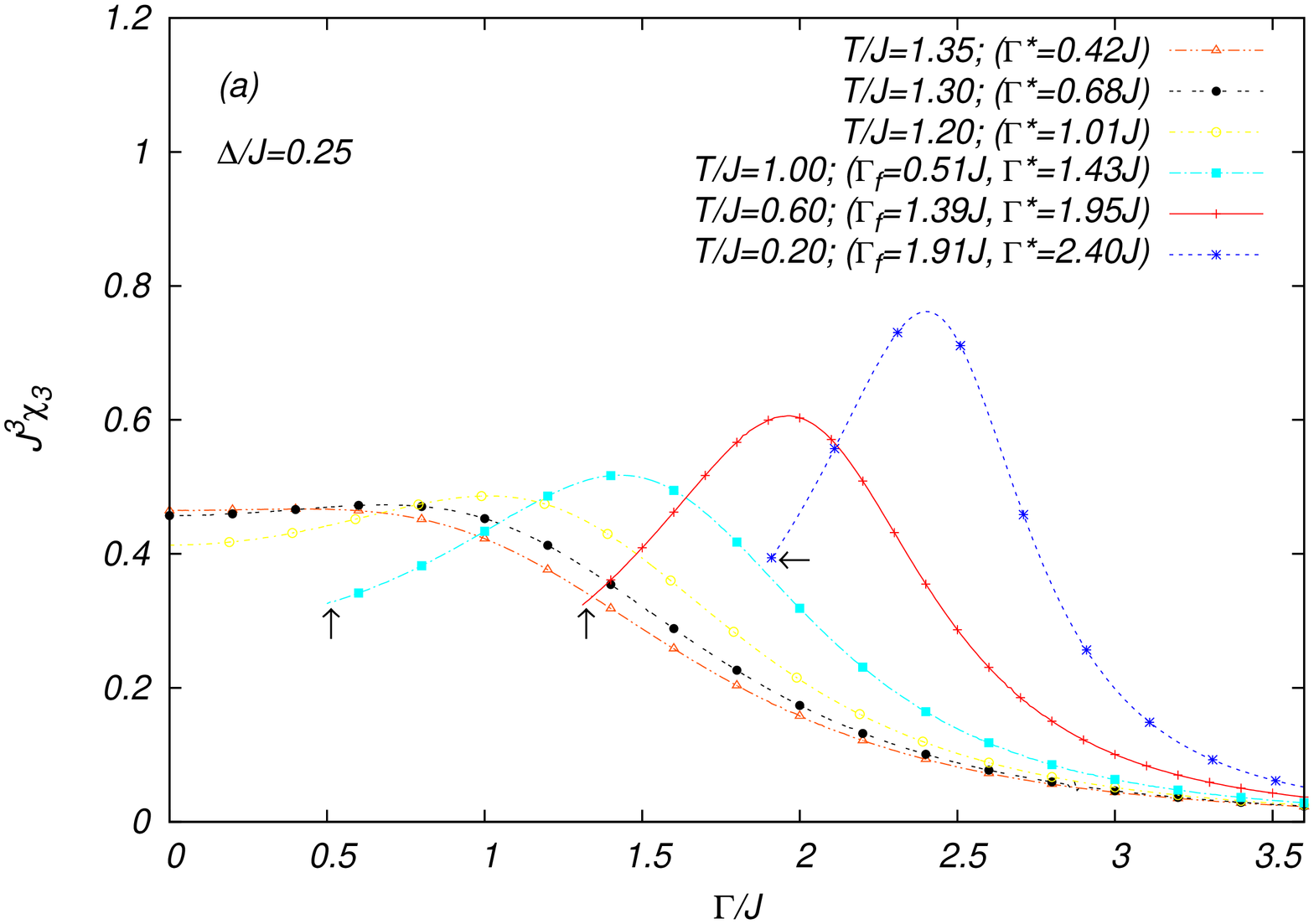}
\includegraphics[width=6.1cm,angle=-90]{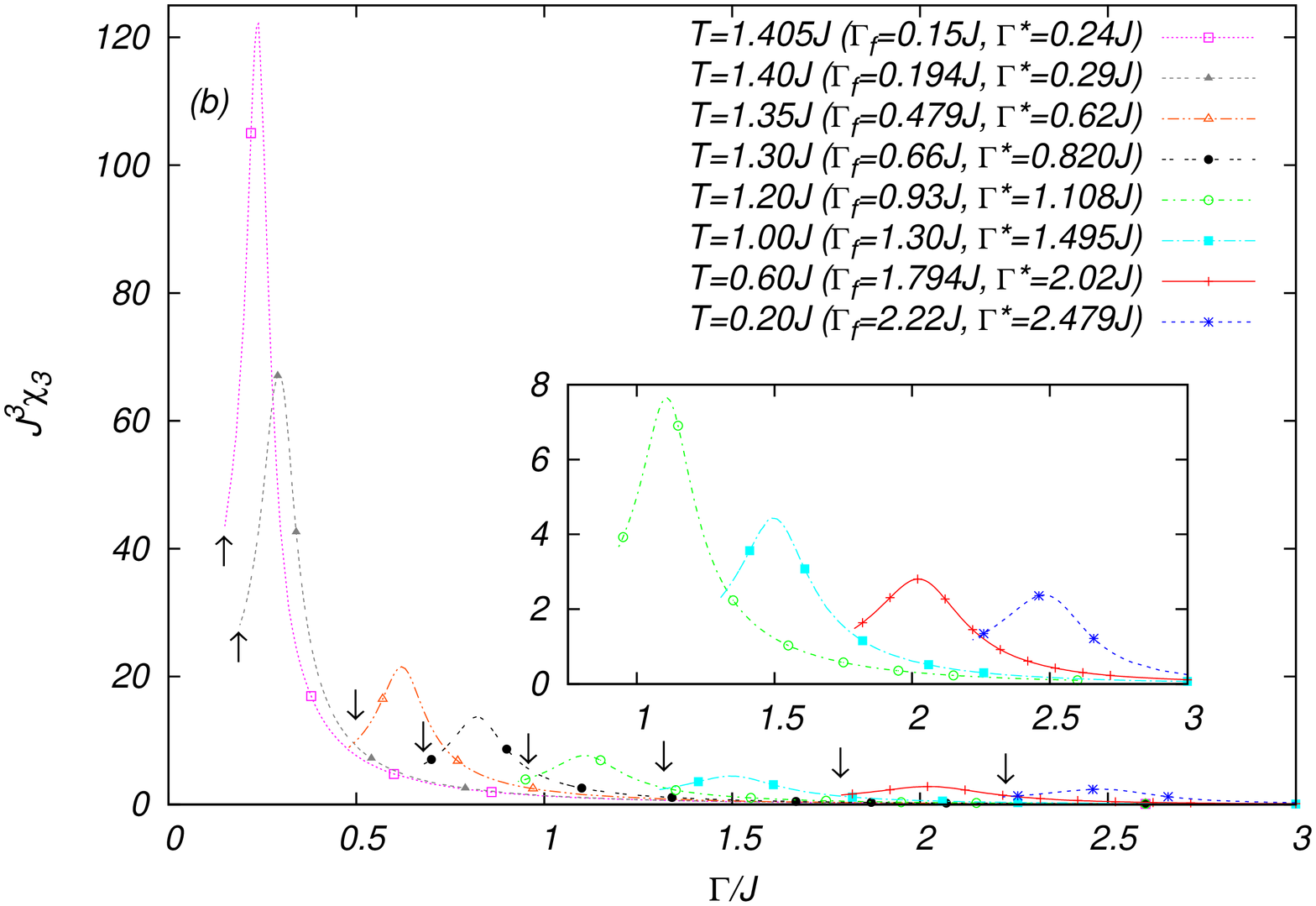}
\caption{ 
(a) The dimensionless nonlinear susceptibility is represented 
versus $\Gamma/J$, for typical 
fixed temperatures and a nonzero width for the 
random fields [$(\Delta/J)=0.25$]. 
The divergences that occur for $\Delta=0$ at $\Gamma_f(T)$ 
[following Eq.~(\ref{x3critic})], signalled by arrows in some cases,  
get smoothened due to the random fields, so that their 
corresponding maxima [located at $\Gamma^{*}(T)$] are shifted 
towards higher values of the transverse field,  
i.e., $\Gamma^{*}(T)>\Gamma_f(T)$.
(b) The behavior of the dimensionless nonlinear susceptibility 
is shown versus $\Gamma/J$, for typical fixed temperatures,
by considering a particular relation involving 
$\Delta$ and $\Gamma$
[$(\Delta/J)= 0.02 (\Gamma/J)^{2}$]; the inset represents an 
amplification of the region for higher values of $\Gamma/J$.
In all cases, the maxima [located at $\Gamma^{*}(T)$] appear 
shifted with respect to the onset of RSB [located at $\Gamma_f(T)$,
signalled by arrows in some cases]
towards higher values of the transverse field,  
i.e., $\Gamma^{*}(T)>\Gamma_f(T)$. 
}
\label{fig4}
\end{figure*}
%%%%%%%%%%%%%%%%%%%%%%%%%

In the present problem, clear phase transitions may be verified only for $\Delta=0$, 
like those exhibited in Fig.~\ref{fig2}. 
From the $\chi_{3}$ plots of Fig.~\ref{fig2}(a) one confirms two 
important features shown in Fig.~\ref{fig1}(a), concerning the 
behavior of the order parameters $q_0$, $q_1$ and $\delta$:  
(i) The freezing temperature $T_{f}$, signaled by the divergence of $\chi_{3}$ in 
Fig.~\ref{fig2}(a), coincides with the onset of RSB indicated by the 
parameter $\delta$ of Fig.~\ref{fig1}(a);
(ii) The temperature $T_{f}$ is lowered by increasing values of $\Gamma/J$.  
The critical exponents associated with the divergences of 
$\chi_{3}$ may be obtained by log-log plots, as shown in the insets of  
Fig.~\ref{fig2}. 
In the inset of Fig.~\ref{fig2}(a) we have verified that the behavior
of Eq.~(\ref{x3critic1}) (represented by the dashed-dotted line) fits well the region 
$0.001<(T-T_f)/T_f<0.1$, 
with the same critical exponent, $\gamma=1$, for both values $\Gamma/J=0.0$ 
(full line) and $\Gamma/J=1.0$ (dotted line), 
suggesting that the transverse field $\Gamma$ should not change the universality 
class of the exponent $\gamma$. It is important to mention that this estimate coincides 
with the well-known value found for the SK model~\cite{binderyoung}.
In Fig.~\ref{fig2}(b) one sees divergences of $\chi_3$ at given values 
of $\Gamma$ [defined as $\Gamma_f(T)$ in Eq.~(\ref{x3critic})], for 
two typical fixed temperatures; like in Fig.~\ref{fig2}(a), these divergences 
coincide with the onset of RSB indicated by the 
parameter $\delta$. 
One notices that, as one approaches
zero temperature [cf., e.g., the case $(T/J)=0.2$], the divergence at 
$\Gamma_f(T)$ approaches
the one that occurs at the QCP, $\Gamma_c^0=2\sqrt{2}J$~\cite{Sachdev95}. 
In the inset of Fig.~\ref{fig2}(b), the critical behavior described by Eq.~(\ref{x3critic}) 
(represented by the 
dashed-dotted line) was fulfilled in both cases, showing a good agreement 
in the region $0.001<(\Gamma-\Gamma_f(T))/\Gamma_f(T)<0.1$, with the same  
exponent $\delta^{\prime}=1$ for the two values of temperatures investigated, 
$(T/J)=0.2$ (full line) and $(T/J)=1.0$ (dashed line). 
Hence, similarly to the results of Fig.~\ref{fig2}(a) concerning the critical exponent $\gamma$ 
of Eq.~(\ref{x3critic1}), the present estimates of $\delta^{\prime}$ suggest that the temperature
should not change the universality class of this later exponent.

In agreement with the previous study of the SK model in the 
presence of a Gaussian random field~\cite{CAVJ16}, the smoothening of 
$\chi_3$ is verified in Fig.~\ref{fig3} for the cases $(\Delta/J)>0$.
For instance, Fig.~\ref{fig3}(a) displays $\chi_3$ versus $T/J$, for increasing values 
of $\Delta/J$, in the case $\Gamma=0$, showing that the divergent peak of the 
nonlinear susceptibility is 
replaced by a broad maximum at a temperature $T^*$. 
One observes that such a peak becomes
smoother, decreasing its height for increasing values of $\Delta/J$. 
Particularly, in the inset of Fig.~\ref{fig3}(a) one sees that the temperature range  
$0.001<(T-T^*)/T^*<0.1$ no longer can be fitted by Eq.~(\ref{x3critic1}) with the  
critical exponent $\gamma=1.0$, in the cases $(\Delta/J) > 0$. 
In a similar way, Fig.~\ref{fig3}(b) shows
$\chi_3$ versus $\Gamma/J$, for $(T/J)=1.0$, considering the 
same values for $\Delta/J$ of Fig.~\ref{fig3}(a); again, 
the peak of nonlinear susceptibility gets flattened due to  
the presence of an applied random field, now 
displaying a maximum at $\Gamma^{*}$. 
Consequently, in such cases the 
region $0.001<(\Gamma-\Gamma^{*})/\Gamma^{*}<0.1$
cannot  be fitted by Eq. (\ref{x3critic}) with a critical exponent $\delta^{\prime}=1.0$,
as shown in the inset of Fig.~\ref{fig3}(b).

%%%%%%%%%%%%%%%%%%%%%%%%%  
\begin{figure*}[htb]
\includegraphics[width=6cm,angle=-90]{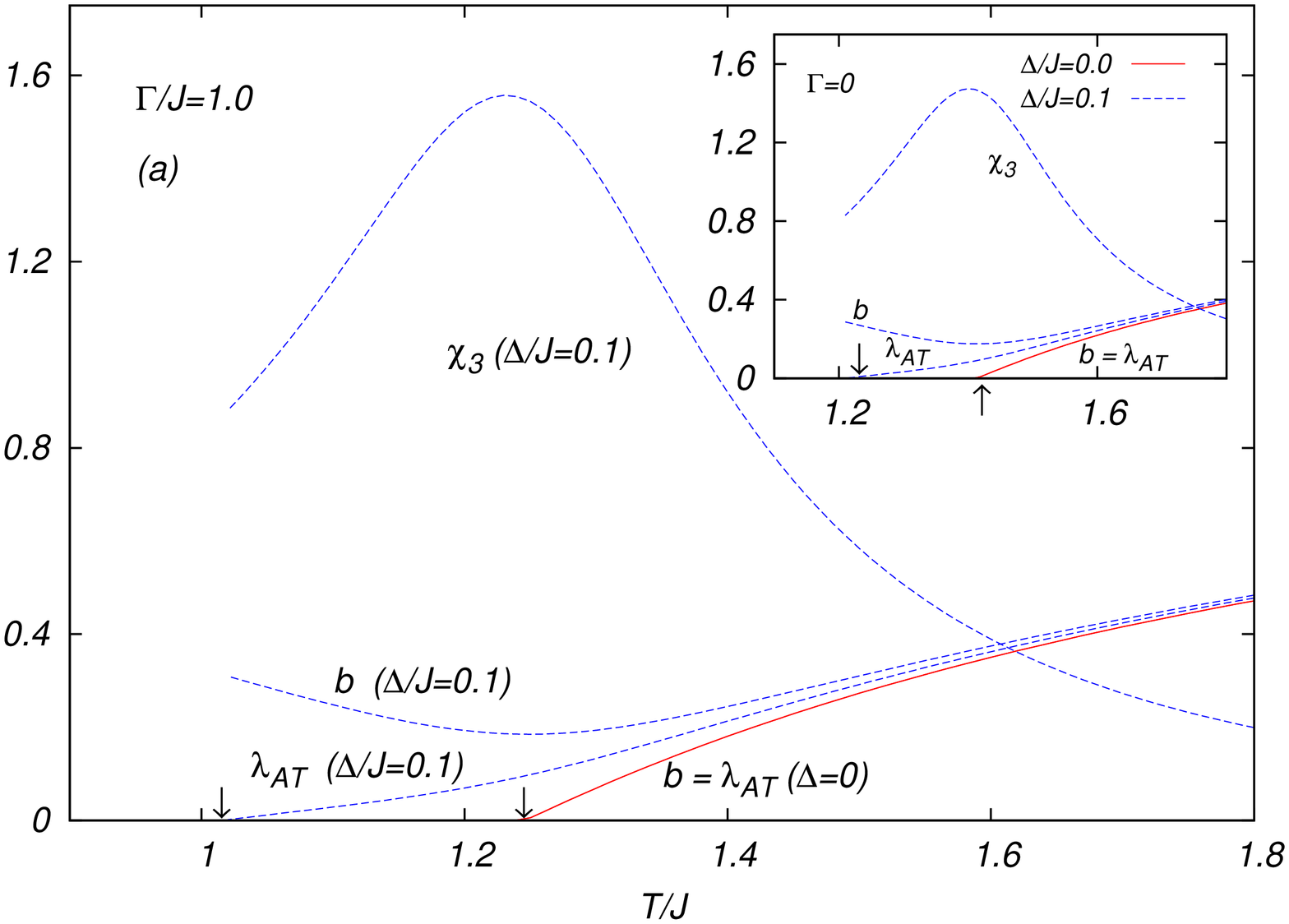} 
\includegraphics[width=6cm,angle=-90]{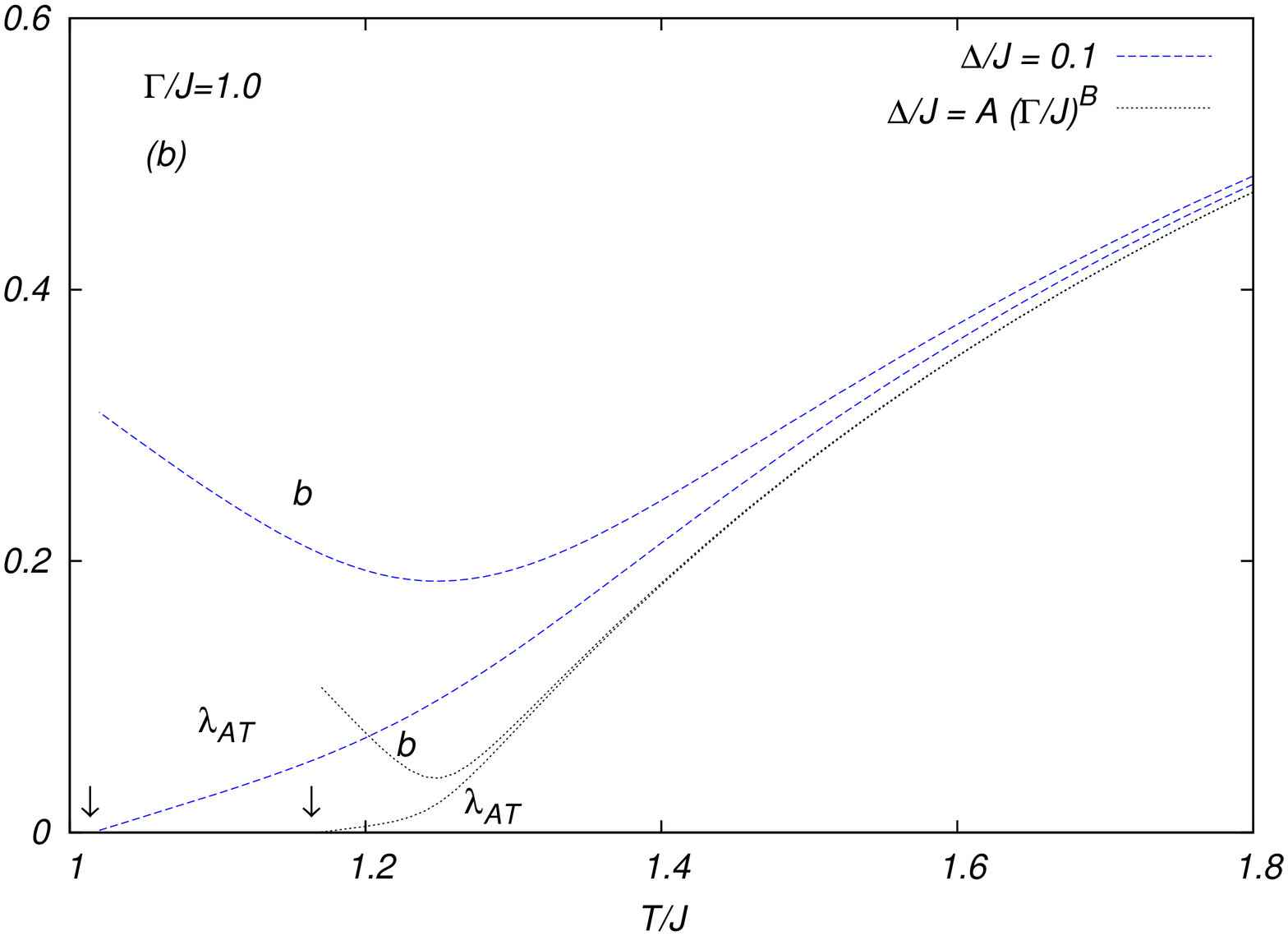} 
\caption{ 
The softening of the nonlinear susceptibility is illustrated by means of the
denominator of  $q_{2}$, i.e., $q_{2} \propto b^{-1}$, 
$b= 1-2(\beta J)^2 I_0(\Gamma)$ [cf. Eq.~(\ref{eq:brsq2})], 
which appears in the expression of $\chi_{3}$
calculated in Appendix A, within the RS approximation.  
The Almeida-Thouless eigenvalue $\lambda_{\rm AT}$, associated with the onset of RSB and 
defining the SG critical temperature $T_{f}$ is also shown, for comparison. 
(a) Results for $(\Gamma/J)=1.0$ are exhibited for two typical values
of $\Delta/J$, namely, $(\Delta/J)=0.0$ and $(\Delta/J)=0.1$, in the case
where $\Gamma$ and $\Delta$ are independent; similar results 
are presented in the inset for $(\Gamma/J)=0$. The full lines 
represent the cases $(\Delta/J)=0.0$,
showing that $\lambda_{\rm AT}$ and the denominator $b$ coincide, becoming 
zero at the temperature $T_{f}$.  
The cases $(\Delta/J)=0.1$ show that $b$ is always positive, 
presenting a smooth minimum around a temperature $T^{*}$, 
leading to the rounding of $\chi_{3}$,
whereas $\lambda_{\rm AT}$ becomes zero at a lower temperature $T_{f}$.  
(b) Results for $(\Gamma/J)=1.0$ and $(\Delta/J)=0.1$ are shown, 
by comparing the case where these two quantities are considered
as independent 
(dashed lines), with the one where they follow the relation proposed in
in in Fig.~\ref{fig4}(b) [$(\Delta/J)= 0.02 (\Gamma/J)^{2}$] (dotted lines). 
In all cases, the arrows locate the freezing temperature $T_f$. 
}
\label{fig5}
\end{figure*}
%%%%%%%%%%%%%%%%%%%%%%%%% 

In Fig.~\ref{fig4} we represent the dimensionless nonlinear 
susceptibility $\chi_{3}$ versus $\Gamma/J$, for typical  
fixed temperatures, in two cases: (a) $\Gamma/J$
and $\Delta/J$ as independent quantities
[Fig.~\ref{fig4}(a)];  
(b) Imposing a relation involving $\Delta$ and
$\Gamma$ [Fig.~\ref{fig4}(b)]. 
In Fig.~\ref{fig4}(a) we consider a fixed value for the width of the 
Gaussian random fields [$(\Delta/J)=0.25$], showing that 
the sharp SG transitions occurring for $\Delta=0$, 
signaled by divergences of $\chi_{3}$
at the corresponding critical
values $\Gamma_f(T)$ [according 
to Eq.~(\ref{x3critic}) and shown by arrows in some curves], change into 
smooth curves with maxima at $\Gamma^{*}(T)$, shifted to higher values 
of $\Gamma$, i.e., $\Gamma^{*}(T)>\Gamma_f(T)$.
Following the proposal of Ref.~\cite{Gingras06}, for dealing properly 
with the experimental behavior 
of $\chi_3$ in the LiHo$_x$Y$_{1-x}$F$_4$ compound, we analyzed 
the present system by imposing a relation involving $\Delta$ and
$\Gamma$, i.e., $\Delta \equiv \Delta(\Gamma)$. 
According to the experimental investigations of Ref.~\cite{Wu93}, such a 
relation should satisfy certain 
requirements, e.g., $\Delta$ should increase monotonically with $\Gamma$,
and one should get $\Delta=0$ for $\Gamma=0$. The simplest proposal 
obeying these conditions comes to be a power function, 
$(\Delta/J)= A (\Gamma/J)^{B}$, where $A$ and $B$ are fitting parameters. 
Herein, these parameters were computed by 
adjusting our results to those of the experiments of Ref.~\cite{Wu93}, 
leading to the optimal values $A=0.02$ and $B=2$.  
In Fig.~\ref{fig4}(b) we exhibit the dimensionless nonlinear susceptibility,     
versus $\Gamma/J$, for typical fixed temperatures,  
by considering this particular relation involving $\Delta$ and $\Gamma$.  
In all cases, the maxima [located at $\Gamma^{*}(T)$] appear 
shifted with respect to the onset of RSB [located at $\Gamma_f(T)$]
towards higher values of the transverse field,  
i.e., $\Gamma^{*}(T)>\Gamma_f(T)$. 
The most important novelty of Fig.~\ref{fig4}(b) [to be contrasted with the results of
Fig.~\ref{fig4}(a)], concerns the fact that the amplitude of the maximum of $\chi_3$
decreases for increasing values of $\Gamma/J$, and consequently, for decreasing
temperatures. 

Recent studies in the compound LiHo$_x$Y$_{1-x}$F$_4$ suggested that
the transverse field $\Gamma$ introduced in the Hamiltonian of Eq.~(\ref{eq1})
should be related to the experimental applied field in a real system,
$H_t$~\cite{Gingras11,Quiliam12};  
in fact, at least for low $H_t$,
$\Gamma \propto H_t^{2}$ (see, e.g., Ref.~\cite{Wu1991}). 
Hence, considering a new dimensionless variable, 
$H_t$ ($H_t \equiv \sqrt{\Gamma/J}$), we have verified that 
the same qualitative behavior shown in both 
Figs.~\ref{fig4}(a) and~\ref{fig4}(b) occur in representations 
of the dimensionless nonlinear 
susceptibility $\chi_{3}$ versus $H_t$.
Consequently, Fig.~\ref{fig4}(b) preserves the agreement with experimental 
observations, showing that besides the progressive smearing of $\chi_3$, 
the amplitude of its maximum also decreases as the real field $H_t$ increases, 
as suggested by Tabei and collaborators \cite{Gingras06}. 

In Appendix A we have calculated $\chi_{3}$ analytically, within the RS
approximation, for both $\Gamma>0$
[cf. Eqs.~(\ref{eq:rschi3}) and~(\ref{eq:rsq2})] and 
$\Gamma=0$ [cf. Eq.~(\ref{x3repl})]. In these calculations, an important
quantity emerged, given in Eq.~(\ref{eq:rsq2}) for $\Gamma>0$, as
\begin{equation}
q_2 = \frac{2 (\beta J)^2 I_0(\Gamma)}{b}~;
\quad b=1-2(\beta J)^2 I_0(\Gamma)~.  
\label{eq:brsq2}
\end{equation}
Notice that the denominator $b$ may become zero, 
leading to a divergence in the nonlinear susceptibility; 
it should be mentioned that $q_{2}$ is the 
only quantity appearing in $\chi_{3}$ [either in  Eq.~(\ref{eq:rschi3}), 
or in Eq.~(\ref{x3repl})],  
which may present a divergence at finite temperatures.  
Moreover, one can show that for $\Delta=0$, the so-called ``dangerous''
eigenvalue~\cite{Almeida78} in Eq.~(\ref{eq:lambdaat}) is equal to the 
denominator of $q_{2}$, i.e., 
$\lambda_{\rm AT} = 1-2(\beta J)^2 I_0(\Gamma)$, 
even for $\Gamma>0$.   
The mechanism behind the flattening of the $\chi_3$ peak at $T^*$ 
is illustrated in Fig.~\ref{fig5}, where we plot the 
quantity $b$ of Eq.~(\ref{eq:brsq2}), $\lambda_{\rm AT}$, 
and $\chi_{3}$, versus the dimensionless temperature, for 
typical choices of $\Gamma/J$ and $\Delta/J$.  
Results for $\Gamma$ and $\Delta$ independent are presented 
in Fig.~\ref{fig5}(a); 
the full lines [cases $(\Delta/J)=0.0$] show that the quantities $b$  
and $\lambda_{\rm AT}$ become zero together, being  
associated with the divergence of $\chi_3$ 
[according to Eq.~(\ref{x3critic1})], signalling the SG phase-transition 
temperature $T_{f}$.  
However, the results for $(\Delta/J)=0.1$ show that such a small value 
for the width of the RFs distribution yields $b>0$, 
which presents a smooth minimum around a temperature $T^{*}$, 
being directly associated with
the rounding behavior of $\chi_{3}$; on the other hand, one has 
$\lambda_{\rm AT}=0$ at a temperature $T_{f}$, with $T_{f}<T^{*}$.   
In Fig. \ref{fig5}(b) we present the denominator $b$ and 
$\lambda_{\rm AT}$, for the cases where $\Gamma$ and $\Delta$ 
are independent (dashed lines), and where these quantities are related 
through the power law
$(\Delta/J)= 0.02 (\Gamma/J)^{2}$. In this later case, since one has  
$(\Delta/J)>0$ for any $(\Gamma/J)>0$, the denominator $b$
will always display a minimum value around a temperature $T^{*}$,
higher than $T_f$. 
Particularly, by means of this relation, higher values of $\Gamma$ imply 
on higher values of $\Delta$, increasing the values of $b$ at 
the minima, resulting in a decrease in the amplitude of the maxima 
of $\chi_3$. 

%%%%%%%%%%%%%%%%%%%%%%%%%
\begin{figure*}[htb]
\includegraphics[width=6cm,angle=-90]{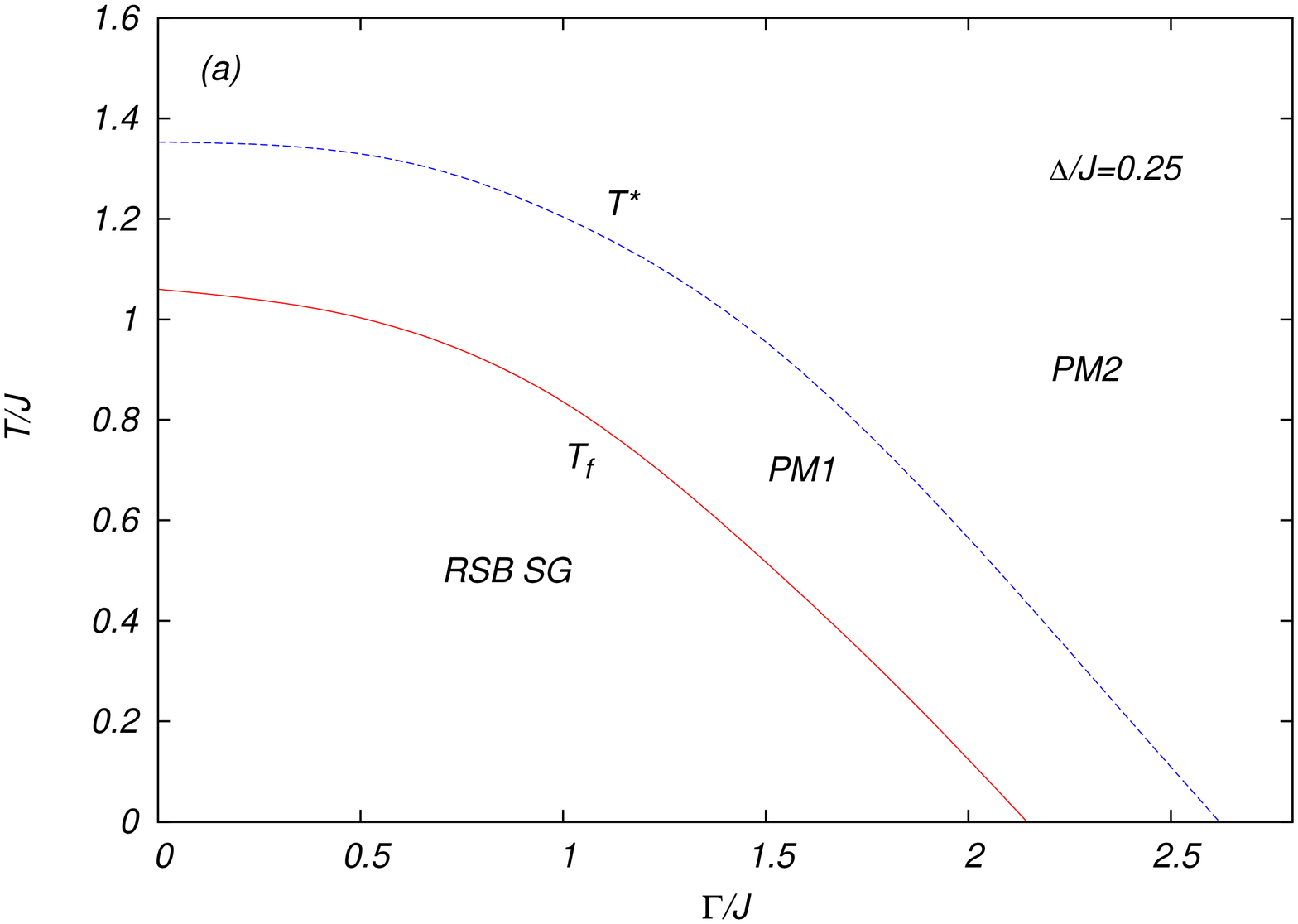} 
\includegraphics[width=6cm,angle=-90]{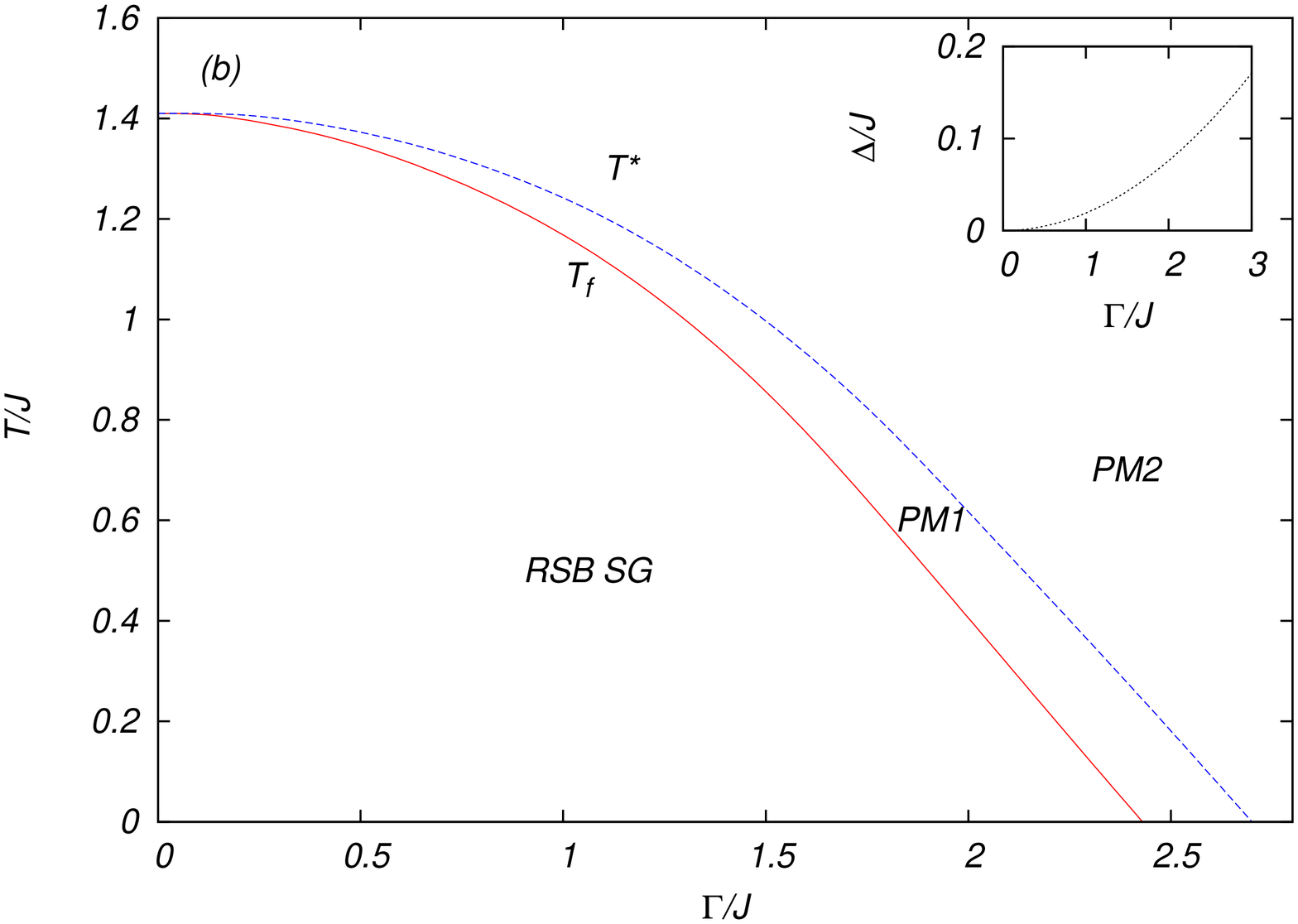} 
\caption{
Phase diagrams $T/J$ versus $\Gamma/J$ showing the frontiers 
separating the SG and paramagnetic phases (full lines), which  
represent the behavior of the temperature $T_{f}$ for increasing 
values of $\Gamma$, located by the onset of RSB, i.e.,
$\lambda_{\rm AT}=0$. 
The dashed lines correspond to the temperature $T^{*}$, associated
with the maximum of the nonlinear susceptibility $\chi_{3}$, and
herein interpreted as a crossover between two distinct 
regions of the paramagnetic phase (PM1 and PM2). 
(a) Phase diagram for $(\Delta/J)=0.25$, in the case
where $\Gamma$ and $\Delta$ are independent. 
(b) Phase diagram for which $\Gamma$ and $\Delta$ 
follow the relation proposed
in Fig.~\ref{fig4}(b) [$(\Delta/J)= 0.02 (\Gamma/J)^{2}$], 
whose parabolic behavior is presented in the inset.  
Due to the usual numerical difficulties, the 
low-temperature results [typically $(T/J)<0.05$)] correspond to 
smooth extrapolations from higher-temperature data.   
}
\label{fig6}
\end{figure*} 
%%%%%%%%%%%%%%%%%%%%%%%%%

In Fig.~\ref{fig6} we present phase diagrams 
$T/J$ versus $\Gamma/J$ showing a decrease in the
temperature $T_{f}$ for increasing 
values of $\Gamma$ (full lines). These lines
delimit the SG phase and were identified with the
onset of RSB, by setting $\lambda_{\rm AT}=0$;
throughout the whole SG phases ones has $\lambda_{\rm AT}<0$. 
The temperature 
$T^{*}$ (dashed lines), associated with the maximum of the nonlinear 
susceptibility $\chi_{3}$, signals a crossover between two 
regions of the paramagnetic phase (PM1 and PM2), as will
be discussed next. 
The phase diagram shown in Fig.~\ref{fig6}(a) corresponds to a 
fixed value of $\Delta$ [$(\Delta/J)=0.25$], and was obtained 
by considering $\Gamma$ and $\Delta$ as independent quantities. 
In this case, one notices that the two lines (full and dashed lines)
remain essentially parallel to one another, up to zero temperature,
where the full line reaches a QCP, which appears to be
shifted towards lower values of $\Gamma$, when
compared with the QCP for $\Delta=0$, 
$\Gamma_c^{0}=2\sqrt{2} J \approx 2.828J$~\cite{zimmer06}. 
The case shown in Fig.~\ref{fig6}(b) corresponds to 
$\Gamma$ and $\Delta$ following the relation 
$(\Delta/J)= 0.02 (\Gamma/J)^{2}$ (see inset), so that
for $\Gamma=0$, one has $\Delta=0$, giving 
$T^{*}=T_{f}$. By increasing values of $\Gamma$, the
width of RFs also increases, leading to a rounded peak in
the nonlinear susceptibility, yielding $T^{*}>T_{f}$, and 
consequently, the region PM1 emerges. Due to 
the joint increase of both $\Gamma$ and $\Delta$, as shown in 
the inset, the region PM1 gets enlarged up to zero temperature, 
where one gets a QCP, shifted towards lower values 
of $\Gamma$ as compared with the QCP for $\Delta=0$, 
similarly to the one occurring in Fig.~\ref{fig6}(a). 

It should be emphasized that the temperature $T^{*}$ plays a role
different from $T_{f}$, in the sense that it does not correspond to
a phase transition, but rather to a crossover between two distinct
regions of the paramagnetic phase. 
The previous analysis of the SK model in the presence of a Gaussian
random field Ref.~\cite{CAVJ16}, which should correspond herein to 
the region of high temperatures and low $\Gamma$ (i.e., the classical
regime), has also found a temperature $T^{*}$, associated with the 
rounded maximum of $\chi_3$, with $T^{*}>T_{f}$.   
In this case, $T^{*}$ was interpreted 
as an effect of the RFs acting inside the paramagnetic phase, instead of some
type of non-trivial ergodicity breaking. Herein, we claim that
the temperature $T^{*}$, although it may be also affected by 
the transverse field $\Gamma$, should be interpreted in a similar 
manner. Hence, along the line signaled by $T_{f}$, the growth of 
$\Gamma$ produces an enhancement of quantum fluctuations, 
which become increasingly dominant as compared with thermal 
fluctuations, driving the 
non-trivial ergodicity breaking of the SG phase transition to a QCP. For temperatures in the region
$T_{f}<T<T^{*}$,  the enhancement of quantum fluctuations by $\Gamma$ along with the spin fluctuations 
due to the RFs inside the paramagnetic phase create two distinct scenarios,  more precisely concerning the PM1 region, as discussed next: 
(a) For fixed $\Delta$ [e.g., Fig.~\ref{fig6}(a)], one
has $\Gamma$ and $\Delta$ independent, so that
the smearing of the nonlinear susceptibility is caused 
only by the RFs, leading to the effect that the full 
and dashed lines remain essentially parallel to one another, 
up to zero temperature;
(b) The phase diagram of Fig.~\ref{fig6}(b), for which 
$\Delta$ and $\Gamma$ are related through the parabolic
behavior shown in the inset, the appearance of $T^{*}$ occurs 
for $\Gamma>0$ (i.e., $\Delta>0$).
Hence, the region PM1 starts very narrow for low $\Gamma$,
and gets enlarged for increasing values of $\Gamma$,
showing that the rounding of $\chi_3$ is dominated by the
enhancement of the RFs, leading to spins fluctuations due to 
the RFs inside the paramagnetic phase.  

\section{Conclusions}

We have investigated a quantum spin-glass model in the presence of a uniform 
transverse field $\Gamma$, as well as of a longitudinal random field $h_{i}$, 
the later following a Gaussian distribution characterized by a width 
proportional to $\Delta$. The model was considered in the limit of 
infinite-range interactions and studied through the replica formalism, 
within a one-step replica-symmetry-breaking procedure. The spin-glass
critical frontier, signaled by the temperature $T_{f}$, 
was identified with the onset of replica-symmetry 
breaking, calculated through the Almeida-Thouless eigenvalue (replicon)
$\lambda_{\rm AT}$, i.e., by setting $\lambda_{\rm AT}=0$. In this approach,
the whole spin-glass phase becomes characterized by $\lambda_{\rm AT}<0$, and 
consequently, it was treated through replica-symmetry breaking.  
Such analysis was motivated 
by experimental investigations on the  
LiHo$_x$Y$_{1-x}$F$_4$ compound. In this system,    
the application of a transverse magnetic field yields
rather intriguing effects, particularly related to the behavior
of the nonlinear magnetic susceptibility $\chi_{3}$, which have 
led to a considerable experimental and theoretical debate. 

We have analyzed two physically distinct situations, namely,
$\Delta$ and $\Gamma$ considered as independent, as well as 
these two quantities related, as proposed recently by some 
authors (see, e.g., Ref.~\cite{Gingras06}). 
In both cases, we have found a spin-glass critical frontier, given 
by $T_{f} \equiv T_{f}(\Gamma, \Delta)$, with such phase being characterized by a 
nontrivial ergodicity breaking.
In the first case, for $\Delta$ fixed, we have found that $T_{f}(\Gamma, \Delta)$ decreases
by increasing $\Gamma$ towards a quantum critical point at
zero temperature, whereas in the second, we have found a similar behavior for this 
critical frontier, with $\Delta$ changing according to variations in $\Gamma$.   
In this later case, we have taken into account  previous experimental investigations~\cite{Wu93}
which suggest that a relation of the type $\Delta \equiv \Delta(\Gamma)$ should satisfy certain 
requirements, e.g., $\Delta$ should increase monotonically with $\Gamma$,
and one should get $\Delta=0$ for $\Gamma=0$. Although such a relation
may not be unique, the simplest proposal following such conditions
appears to be a power function,   
$(\Delta/J)= A (\Gamma/J)^{B}$. 
In the present work, the parameters $A$ and $B$ were computed by 
adjusting our results to those of the experiments of Ref.~\cite{Wu93}, 
leading to the  
the optimal values $A=0.02$ and $B=2$.  

We have shown that the present approach, considering the 
relation $(\Delta/J)= 0.02 (\Gamma/J)^{2}$, was able to reproduce 
adequately the experimental observations on the 
LiHo$_x$Y$_{1-x}$F$_4$ compound, with theoretical results
coinciding qualitatively 
with measurements of the nonlinear susceptibility $\chi_{3}$.
As a consequence, by increasing $\Gamma$ gradually, our results indicate 
that $\chi_{3}$ becomes progressively rounded, presenting a maximum
at a temperature $T^{*}$ ($T^{*}>T_{f}$); moreover, both 
amplitude of the maximum and the value of $T^{*}$
diminish, by enhancing $\Gamma$. 

From the analysis where $\Delta$ and $\Gamma$ are considered as
independent, we have concluded that the random field is the main responsible 
for the smearing of the nonlinear susceptibility. Hence, the random
field acts significantly  
inside the paramagnetic phase, leading to two regimes delimited
by the temperature $T^{*}$, 
one for $T_{f}<T<T^{*}$ (called herein as PM1), and another one 
for $T>T^{*}$ (denominated as PM2). 
In the paramagnetic regime for $T>T^{*}$ 
one should have weak correlations and consequently, the usual paramagnetic type of 
behavior. However, close to  $T^{*}$, and particularly for temperatures
in the range $T_{f}<T<T^{*}$, one expects a rather nontrivial 
behavior in real systems, as happens with experiments in the compound 
LiHo$_x$Y$_{1-x}$F$_4$, resulting in very controversial 
interpretations~\cite{Schechter1,Laflorencie06,Gingras06,Gingras08,Gingras11,%
Jonsson07,Torres08,Mydosh15}.
Hence, as already argued in the analysis of the SK model in the presence of 
Gaussian random field~\cite{CAVJ16}, the line PM1--PM2
may not characterize a real phase 
transition, in the sense of a diverging $\chi_{3}$, but the region 
PM1 should be certainly characterized by a rather nontrivial dynamics.
As one possibility, one should have a growth of free-energy barriers 
in this region, leading to a slow dynamics, 
whereas only below $T_{f}$ the nontrivial ergodicity breaking appears, 
typical of RSB in SG systems.  Also, one could have 
Griffiths singularities along PM1, which are found currently
in disordered magnetic systems, like site-diluted 
ferromagnets~\cite{griffithssing}, ferromagnet in a random
field~\cite{dotsenkosing},
classical Ising spin glasses~\cite{randeria1985}, 
and also claimed to occur
in quantum spin glasses~\cite{Guo96,Rieger96,Young96}. 
Whether such curious properties may appear throughout the 
region PM1 in the present problem, represents a matter for further 
investigation. 
In fact, recent experiments in the above compound
for $x=0.045$ strongly suggest this picture~\cite{biltmo2012}: 
these authors  claim an ``unreachable'' transition due to an ultra-slow dynamics
(of the order $10^{7}$ times slower than the ones of
conventional spin-glass materials) and argue 
that such a dynamics should be caused by a Griffiths phase between the 
paramagnetic and spin-glass phases.  

Next, we discuss some contributions of the present work, as compared to
previous theoretical approaches in this problem.
(i) The analysis of Ref.~\cite{Kim02} did not take into account the random 
field, which in our view, represents a key ingredient for an appropriate description 
of the properties of LiHo$_x$Y$_{1-x}$F$_4$. Moreover, as it was
shown herein, the RSB SG parameters, together with the magnetization $m$, and
the quadrupolar parameter $p$, all  form a set 
of coupled equations, to be solved simultaneously. The approach of Ref.~\cite{Kim02}
considered $p$ as independent from the remaining parameters; this could be directly 
related with the curious result concerning a part of the SG phase 
characterized by stability of the 
replica-symmetric solution, along which these authors find the 
rounded maximum of $\chi_3$.  
(ii) The study of Ref.~\cite{Gingras06} has considered an effective Hamiltonian
characterized by an extra two-body interacting term (as compared with the Hamiltonian 
used herein), coupling spin operators in the $x$ and $z$ directions.
Moreover, these authors have suggested a relation $\Delta \equiv \Delta(\Gamma)$,
which due to the Hamiltonian employed, turned out 
to be slightly different from ours, e.g., $(\Delta/J)= A (\Gamma/J)^{B}$, with an 
exponent $B<1$. The results obtained herein for the nonlinear susceptibility $\chi_{3}$
corroborate those of Ref.~\cite{Gingras06}; 
however, we understand that the present analysis, characterized by a single 
two-body interacting term, $- \sum_{(i,j)} J_{ij} \hat S_{i}^{z} \hat S_{j}^{z}$,
leads to a much simpler analysis to the problem, when compared to the one 
carried in this previous work.

To conclude, we have considered a model able to reproduce theoretically many properties
observed in experiments on the LiHo$_x$Y$_{1-x}$F$_4$ compound, particularly those
related to the nonlinear susceptibility $\chi_{3}$.  The present
theoretical proposal appears to be simpler than previous ones, and consequently,
its results should be easier to compare with further experimental investigations. 
Obviously, the observation of a clear spin-glass state, characterized by a nontrivial 
ergodicity breaking at a temperature $T_{f}$ (below the temperature $T^{*}$ where one
observes rounded effects on the nonlinear susceptibility) represents a challenge for
experiments.  

\begin{acknowledgments}
We acknowledge the partial financial support from CNPq, 
FAPERGS, FAPERJ, and CAPES (Brazilian funding agencies). 
\end{acknowledgments}  
  
\appendix

\section{Nonlinear Susceptibility in the RS Solution}

In this appendix we obtain the nonlinear susceptibility $\chi_3$ analytically for the 2S model, 
within the RS solution. Although in  the RS solution, these results 
allow us to analyze in detail how the RFs and the transverse field 
$\Gamma$ affect the nonlinear susceptibility. Particularly, one has that
the nonlinear susceptibility of Eq.~(\ref{chi3}) becomes
\begin{equation}
\begin{split}
\chi_3 &= -\frac{1}{3!} \left. \frac{ \partial^3 m}{ \partial H_l^3} 
\right|_{H_l \rightarrow 0 } \\  
 &= \frac{\beta^3}{3}\left[1 + 3  q_2+ 2(\beta J)^2 V_3\right] \frac{-V_2}{2-2(\beta J)^2 V_3}~,                 
 \end{split}
 \label{eq:rschi3}
 \end{equation}
where
\begin{equation}
q_2=\frac{2 (\beta J)^2 I_0(\Gamma)}{1-2(\beta J)^2 I_0(\Gamma)}~, 
\label{eq:rsq2}
\end{equation}
with the following definitions
\begin{equation}
I_0(\Gamma)=  \frac{V_3-V_2}{2} + 2(\beta J)^2 \frac{V_2 V_1}{2-2(\beta J)^2 V_3}~, 
\end{equation}
\begin{equation}
 V_1=\int Dz \left[\frac{ C_3 C_1}{K^2}
 - \frac{ C_2 ( C_1)^2}{K^3}\right]~, 
\end{equation}
\vskip \baselineskip
\begin{equation}
\begin{split}
V_2&=\int Dz \left[ \frac{ C_4 }{K} \right. - 4 \frac{C_3 C_1}{K^2}
 -3 \left(\frac{C_2}{K}\right)^2 \\
 &+12 \frac{C_2 (C_1)^2}{K^3}
- \left. 6 \left(\frac{ C_1} {K} \right)^4\right]~, 
\end{split}
\end{equation}

\begin{equation}
\begin{split}
V_3=\int Dz \left[ \frac{ C_4}{K} \right. - 2 \frac{ C_3 C_1}{K^2}
- \left(\frac{ C_2}{K}\right)^2 
 +2  \left.  \frac{ C_2 (C_1)^2}{K^3}\right]~.  
\end{split}
\end{equation}
In the equations above, one has that
\begin{equation}
C_n = \int D\xi \frac{\partial^n f(h)}{\partial h^n}~; \quad 
K=  \int D\xi \, f[h(z,\xi),\Gamma]~, 
\end{equation}
with
\begin{equation}
f[h(z,\xi),\Gamma]=\cosh\sqrt{h^{2}(z,\xi)+(\beta \Gamma)^2}~, 
\end{equation}
and 
\begin{equation}
h(z,\xi)=\beta J[\sqrt{2q+(\Delta/J)^2}z+\sqrt{2(p-q)}\xi]~. 
\end{equation}
In addition, the limit of stability of the RS solution is delimited 
by $\lambda_{\rm AT}>0$~\cite{Almeida78}, which may be expressed
in terms of the above quantities as  
\begin{equation}
\lambda_{\rm AT} = 1 - 2(\beta J)^2\int Dz\left[ \frac{C_2}{K} - \left(\frac{ C_1}{K}\right)^2 \right]^2.
\label{eq:lambdaat}
\end{equation}
The particular case $\Gamma=0$ gives 
\begin{equation}
h_0(z) =\beta J(\sqrt{2q+(\Delta/J)^2}z~, 
\end{equation}
as well as $V_1=V_3=0$, whereas 
\begin{equation}
V_2=-2 \int Dz [\mbox{sech}^4 h_0(z) - 2 \tanh^2 h_0(z) \mbox{sech}^2 h_0(z)]~.
\end{equation}
As a result, $\chi_3$ becomes
\begin{equation}
\label{x3repl}
 \chi_3  = \frac{\beta^3}{3}\left[1 + 3  q_2 \right] I_0(0)
 \end{equation}
 where $q_2=[2(\beta J)^2 I_0(0)]/[1-2(\beta J)^2 I_0(0)]$ and 
\begin{equation}
 I_0(0) = \int Dz [\mbox{sech}^4 h_0(z) - 2 \tanh^2 h_0(z) \mbox{sech}^2 h_0(z)].  
\end{equation}
In this case, $\chi_3$ coincides with the one found in Ref.~\cite{CAVJ16}.

\end{document}